\documentclass[12pt]{article}
\usepackage{epsfig, epsf, graphicx, subfigure, color}
\usepackage{pstricks, pst-node, psfrag}
\usepackage{amssymb,amsmath}
\usepackage{verbatim,enumerate}
\usepackage{rotating, lscape}
\usepackage{setspace}
\usepackage{bbm}
\usepackage{natbib}
\usepackage{amsthm}
\usepackage{url}
\usepackage{appendix}
\usepackage{gensymb}

\usepackage{caption}
\usepackage{hyperref}

\hypersetup{
  colorlinks   = true, 
  urlcolor     = blue, 
  linkcolor    = blue, 
  citecolor   = blue 
}

\theoremstyle{definition}

\newcommand{\R}{\mathbb{R}}
\newcommand{\s}{\textbf{s}}
\newcommand{\h}{\textbf{h}}

\theoremstyle{remark}

\newcommand{\argmaxE}{\mathop{\mathrm{argmax}}}          

\newtheorem{theorem}{Theorem}

\def\trans{^{T}}
\def\tr{{\mathrm{tr}}}
\newcommand{\bi}{\begin{itemize}}
	\newcommand{\ei}{\end{itemize}}

\begin{document}
	\thispagestyle{empty} \baselineskip=28pt \vskip 5mm
	\begin{center} {\Large{\bf Modeling and Predicting Spatio-temporal Dynamics of PM$_{2.5}$ Concentrations Through Time-evolving Covariance Models}}
	\end{center}
	
	\baselineskip=12pt \vskip 5mm
	
	\begin{center}\large
		Ghulam A. Qadir\footnote[1]{ Heidelberg Institute for Theoretical Studies, Schloß-Wolfsbrunnenweg 35, 69118 Heidelberg, Germany. E-mail: ghulam.qadir@h-its.org
			\baselineskip=10pt 
		}, and Ying Sun\footnote[2]{
			\baselineskip=10pt CEMSE Division, King Abdullah University of Science and Technology, Thuwal 23955-6900, Saudi Arabia.
			E-mail: ying.sun@kaust.edu.sa}\end{center}
	
	\baselineskip=16pt \vskip 1mm \centerline{\today} \vskip 8mm
	
	\begin{center}
		{\large{\bf Abstract}}\end{center}
        
        Fine particulate matter (PM$_{2.5}$) has become a great concern worldwide due to its adverse health effects. PM$_{2.5}$ concentrations typically exhibit complex spatio-temporal variations. Both the mean and the spatio-temporal dependence evolve with time due to seasonality, which makes the statistical analysis of PM$_{2.5}$ challenging. In geostatistics, Gaussian process is a powerful tool for characterizing and predicting such spatio-temporal dynamics, for which the specification of a spatio-temporal covariance function is the key. While the extant literature offers a wide range of choices for flexible stationary spatio-temporal covariance models, the temporally evolving spatio-temporal dependence has received scant attention only. To this end, we propose a time-varying spatio-temporal covariance model for describing the time-evolving spatio-temporal dependence in PM$_{2.5}$ concentrations. For estimation, we develop a composite likelihood-based procedure to handle large spatio-temporal datasets.The proposed model is shown to outperform traditionally used models through simulation studies in terms of predictions. We apply our model to analyze the PM$_{2.5}$ data in the state of Oregon, US. Therein, we show that the spatial scale and smoothness exhibit periodicity. The proposed model is also shown to be beneficial over traditionally used models on this dataset for predictions.
	
	\baselineskip=17pt

	\begin{doublespace}
		
		\par\vfill\noindent
		{\bf Some key words}: Spatio-temporal covariance, Nonstationarity, Mat\'ern covariance function, Bernstein functions.
	
	\end{doublespace}
	
	\clearpage\pagebreak\newpage \pagenumbering{arabic}
	\baselineskip=26.5pt
	

\def\trans{^{T}}
\def\tr{{\mathrm{tr}}}

\def\edijt{\Sigma_{\mathcal{S}_{ij},\mathcal{T}}}
\def\edtij{\Sigma_{\mathcal{S},\mathcal{T}_{ij}}}
\def\ydijt{\boldsymbol{X}_{\mathcal{S}_{ij},\mathcal{T}}}
\def\ydijdt{\boldsymbol{X}_{\mathcal{S}_{ij'},\mathcal{T}}}

\def\ydtij{\boldsymbol{X}_{\mathcal{S},\mathcal{T}_{ij}}}
\def\ydtkl{\boldsymbol{X}_{\mathcal{S},\mathcal{T}_{kl}}}

\def\ldijtr{{L}^r_{\mathcal{S}_{ij},\mathcal{T}}}
\def\ldijdtr{{L}^r_{\mathcal{S}_{ij'},\mathcal{T}}}

\def\ldtijr{{L}^r_{\mathcal{S},\mathcal{T}_{ij}}}
\def\ldtklr{{L}^r_{\mathcal{S},\mathcal{T}_{kl}}}

\def\ldijts{{L}^s_{\mathcal{S}_{ij},\mathcal{T}}}
\def\ldtijs{{L}^s_{\mathcal{S},\mathcal{T}_{ij}}}

\def\ydljt{\boldsymbol{X}_{\mathcal{S}_{lj},\mathcal{T}}}
\def\ydtlj{\boldsymbol{X}_{\mathcal{S},\mathcal{T}_{lj}}}

\def\ydmjt{\boldsymbol{X}_{\mathcal{S}_{mj},\mathcal{T}}}
\def\ydtmj{\boldsymbol{X}_{\mathcal{S},\mathcal{T}_{mj}}}

\def\ydmnt{\boldsymbol{X}_{\mathcal{S}_{mn},\mathcal{T}}}
\def\ydtmn{\boldsymbol{X}_{\mathcal{S},\mathcal{T}_{mn}}}

\def\ldljtr{{L}^r_{\mathcal{S}_{lj},\mathcal{T}}}
\def\ldtljr{{L}^r_{\mathcal{S},\mathcal{T}_{lj}}}
\def\ldljts{{L}^s_{\mathcal{S}_{lj},\mathcal{T}}}
\def\ldtljs{{L}^s_{\mathcal{S},\mathcal{T}_{lj}}}

\def\ldmjtr{{L}^r_{\mathcal{S}_{mj},\mathcal{T}}}
\def\ldtmjr{{L}^r_{\mathcal{S},\mathcal{T}_{mj}}}
\def\ldmjts{{L}^s_{\mathcal{S}_{mj},\mathcal{T}}}
\def\ldtmjs{{L}^s_{\mathcal{S},\mathcal{T}_{mj}}}

\def\ldmntr{{L}^r_{\mathcal{S}_{mn},\mathcal{T}}}
\def\ldtmnr{{L}^r_{\mathcal{S},\mathcal{T}_{mn}}}
\def\ldmnts{{L}^s_{\mathcal{S}_{mn},\mathcal{T}}}
\def\ldtmns{{L}^s_{\mathcal{S},\mathcal{T}_{mn}}}

\section{Introduction}\label{sec:intro}

In the context of air quality control, particulate matter concentrate with diameter $\leq2.5\mu m$ (PM$_{2.5}$) is a crucial pollutant of concern because of its deleterious effects on human health 
 \citep{dominici,pope,samoli,chang}. In consequence, PM$_{2.5}$ has been a focal topic in numerous air quality control oriented research where it has been studied for its chemical composition \citep{YE2003499,CHENG201596,Zhang_nature2020}, risk assessment \citep{riskasses1,riskasses2}, statistical modeling and  prediction \citep{pmmodelpred0,pm25modelpred1,qadir2020a, qadir2020b}, etc.  PM$_{2.5}$ is closely connected to the meteorology \citep{sheehan2001estimated,dawson2007sensitivity,TAI20103976}, which causes the seasonality or other time-varying factors to have a strong influence on it. This strong seasonality effect has been noted in numerous case studies pertaining to the spatio-temporal variations of PM$_{2.5}$ \citep{chow2006,bell2007,Xiulingzhao2019}. Much of the literature concerning the spatio-temporal modeling of PM$_{2.5}$, such as \cite{seungjaelee2012,li2017ensemble,lichen2018,luxiao2018}, account for such time-varying effects in the mean but ignore those effects in the spatio-temporal covariance of PM$_{2.5}$. In order to achieve a more comprehensive spatio-temporal modeling of PM$_{2.5}$, both the mean and spatio-temporal dependence should be allowed to evolve temporally for the seasonality effect. 
This necessitates the development of flexible time-varying model for the purpose of beneficial statistical modeling and prediction of  PM$_{2.5}$. 

Statistical modeling of PM$_{2.5}$ as a spatio-temporal stochastic process allows us to delve into the associated spatio-temporal uncertainties and perform predictions at unobserved locations and time points, which can be beneficial in planning strategies for air quality control and formulating health care policies. The Gaussian process models are the typical choice of stochastic process models in spatio-temporal modeling where the joint distribution of random variables continuously indexed with space and time is multivariate normal. In particular, let $X(\s,t),\: (\s,t)\in \R^d \times \R$, be the Gaussian process, indexed by space-time coordinates $(\s,t)$, then for any finite set of space-time pairs $\{(\s_1,t_1),\ldots,(\s_{n},t_n)\},\;n\geq 1$, the random vector $\{X(\s_1,t_1),\ldots,X(\s_{n},t_{n})\}^{\text{T}}\sim\mathcal{MVN}(\boldsymbol{\mu}_{n\times 1},\boldsymbol{\Sigma}_{n\times n})$, where $\boldsymbol{\Sigma}=[\text{Cov}\{X(\s_i,t_i),X(\s_j,t_j)\}]_{i,j=1}^n$ is the covariance matrix and $\boldsymbol{\mu}=[\mathbb{E}\{X(\s_1,t_1)\},\ldots,\mathbb{E}\{X(\s_n,t_n)\}]^\text{T}$ is the mean vector of the multivariate normal distribution. The entries of $\boldsymbol{\Sigma}$ are usually defined through some nonnegative definite parametric function $\text{K}(\s_i,\s_j,t_i,t_j)$. 
The optimal prediction of an unobserved part of $X(\s,t)$ is given by kriging predictor \citep{ncressie}, which is a weighted linear combination of the observed part of $X(\s,t)$. These weights are affected by the covariance structure of the process, and therefore, the function $\text{K}(\s_i,\s_j,t_i,t_j)$ must be specified diligently to obtain accurate predictions.

For practical convenience, it is often assumed that the covariance function $\text{K}(\s_i,\s_i+\textbf{h},t_i,t_i+u)$ is stationary in space and time, i.e., $\text{K}(\s_i,\s_i+\textbf{h},t_i,t_i+u)=\text{C}(\textbf{h},u)$ depends only on the spatial lag $\textbf{h}$ and temporal lag $u$. The particular restrictions $\text{C}(\textbf{h},0)$ and $\text{C}(\boldsymbol{0},u)$ represent purely spatial and purely temporal covariance functions, respectively. The existing literature on stationary spatio-temporal models provides practitioners with numerous alternatives, and those are comprehensively summarized in review papers by \cite{kyriakidis}, \cite{gneiting2006geostatistical} and \cite{wchen}. A rudimentary approach to build a valid spatio-temporal covariance function is to impose separabilty, in which $\text{C}(\textbf{h},u)$ can be decomposed into purely spatial and purely temporal covariance function. This decomposition can be in the form of a product: $\text{C}(\textbf{h},u)=\text{C}_s(\textbf{h})\text{C}_t(u)$ (see \cite{rod,de1997spatial2} for application), or in the form of a sum: $\text{C}(\textbf{h},u)=\text{C}_s(\textbf{h})+\text{C}_t(u)$. The sum-based model suffers the problem of rendering singular covariance matrices for some configurations of spatio-temporal data \citep{myer_journel,rouhani}. Besides, a major shortcoming of separable model is its inability to allow for any space-time interactions which are often present in real data. To allow for space-time interactions, \cite{cressie_huang1999} introduced some classes of stationary nonseparable spatio-temporal covariance functions based on Fourier transform pairs in $\R^d$. Their approach led \cite{gneiting2002} to develop general classes of stationary nonseparable spatio-temporal covariance functions which are constructed using completely monotone functions and positive functions with completely monotone derivatives. Some further developments of nonseparable models include \cite{stein2005,DeIaco,fuentes2008}.

Among the existing stationary nonseparable spatio-temporal covariance functions, \cite{gneiting2002}'s classes have been notably popular and are explored for further generalizations \citep{porcu2006,porcu2016}. Specifically, \citeauthor{gneiting2002}'s class is defined as: \begin{equation}\label{eq1}
\text{C}(\h,u)=\frac{\sigma^2}{\psi(|u|^2)^{d/2}}\varphi\Bigg(\frac{\|\h\|^2}{\psi(|u|^2)}\Bigg), (\h,u)\in\R^d\times\R,\end{equation}
where $\sigma>0$ is the standard deviation of the process, $\varphi(w),\:w\geq0,$ be any completely monotone function and $\psi(w),\: w\geq0,$ be any positive function with a completely monotone derivative which is commonly termed as Bernstein function \citep{bhatiajain,porcu2011bern}. Table 1 and Table 2 of \cite{gneiting2002} provide different choices of $\varphi(\cdot)$ and standardized $\psi(\cdot), \psi(0)=1$, respectively. For a particular choice $\varphi(w)=(\alpha w^{1/2})^\nu K_\nu(\alpha w^{1/2})/\{2^{\nu-1}\Gamma(\nu)\},\:$ $ \alpha>0, \nu>0$, where $K_\nu(\cdot)$ denotes a modified Bessel function of the second kind of the order $\nu$ \citep{abramowitz1965handbook}, \eqref{eq1} reduces to: \begin{equation}\label{eq2}
\text{C}(\h,u)=\frac{1}{\psi(|u|^2)^{d/2}}\frac{\sigma^2}{2^{\nu-1}\Gamma(\nu)}\Bigg(\frac{\alpha\|\h\|}{\psi(|u|^2)^{1/2}}\Bigg)^\nu K_\nu \Bigg(\frac{\alpha\|\h\|}{\psi(|u|^2)^{1/2}}\Bigg), (\h,u)\in\R^d\times\R.
\end{equation}  

The purely spatial covariance function in \eqref{eq2}: $\text{C}(\h,0)=\sigma^2(\alpha\|\h\|)^\nu K_\nu(\alpha\|\h\|)2^{1-\nu}/\Gamma(\nu)$, belongs to the Mat\'ern class \citep{matern,guttorp}, henceforth denoted as $\sigma^2\text{M}(\textbf{h}\mid \alpha, \nu)$, where $\alpha>0$ and $\nu>0$ represent spatial scale and smoothness parameters, respectively. The Mat\'ern class has become an extremely preferred and important class of isotropic covariance functions for modeling spatial data \citep{stein2012,bimat,multimat}, and therefore, \eqref{eq2} which we hereafter refer as ``Gneiting-Mat\'ern'' class is also particularly important.

The class of stationary spatio-temporal models which does not allow covariance to evolve either in space or time, can be restrictive for many real applications. 
Therefore, 
while this class of models is essential, its relevance to the considered data must be assessed individually. Moreover, the spatio-temporal data which is likely to demonstrate heterogeneity of dependence (nonstationarity) in space and/or time must be served with flexible space and/or time-varying models for satisfactory inference and prediction. Consequently, numerous nonstationary spatio-temporal models with varied fundamental constructions have been proposed in the last two decades. \cite{stroud2001} proposed a state-space model in which the nonstationarity operates through locally-weighted mixture of regression surfaces with time varying regression coefficients. \cite{ma2002spatio} proposed nonstationary spatio-temporal covariance model constructions through scale and positive power mixtures of stationary covariance functions. Set in the spectral domain, \cite{fuentes2008} derived the nonstationary spatio-temporal covariance model via mixture of locally stationary spatio-temporal spectral densities. \cite{shand2017modeling} extended the idea of dimension expansion by \cite{bornnetal2012} to the spatio-temporal case, resulting in nonstationary spatio-temporal covariances. Some other important works in the nonstationary spatio-temporal modeling include \cite{huang2004,KOLOVOS2004815,bruno2009simple,sigrist,XU2018160}. 
The nonstationary extension of the Gneiting-Mat\'ern class is also of particular interest and have been explored by \cite{porcu2006} and \cite{porcu2007} for spatial anisotropy and spatial nonstationarity, respectively, but the  resulting models in those approaches are stationary in time. Alternatively, one can impart space-time nonstationarity to the Gneiting-Mat\'ern class through the process convolution-based spatio-temporal covariance model of \cite{garg2012learning}, however, their model does not allow for evolving smoothness. 

We propose a time-varying spatio-temporal covariance model by generalizing the Gneiting-Mat\'ern class \eqref{eq2} to include temporally varying spatial scale and smoothness. The foundational idea of the proposed model is analogous to the work of \cite{ip2015time}, however, our model construction significantly differs from that of \cite{ip2015time}. The time-varying model of \cite{ip2015time} requires computing the square root of purely spatial covariance matrices for the evaluation of the full space-time covariance matrix, which can be a computationally expensive evaluation for a large number of spatial locations. In contrast, the proposed model avoids computing the square root of matrices and provides a simple parametric functional form for evaluation of the spatio-temporal covariance. Additionally, since the proposed model is a generalization of the Gneiting-Mat\'ern class, it inherits all the desirable properties of the original class and beyond. 

The rest of the paper is organized as follows: we introduce the considered PM$_{2.5}$ data and perform a preliminary data analysis in Section \ref{eda} to demonstrate the necessity of time-varying model. In Section \ref{method}, we describe the proposed time-varying model, its properties and a composite likelihood-based estimation method. We conduct a simulation study to compare the performance of the proposed time-varying model with Gneiting-Mat\'ern class and separable model in Section \ref{simulation}. The proposed time-varying model is then applied to analyze the PM$_{2.5}$ data in Section \ref{dataanalysis}. We conclude with discussion and potential future extensions in Section \ref{discussion}.

\section{The PM$_{2.5}$ Data and the Preliminary Analysis}\label{eda}

\begin{figure}
\centering     
\includegraphics[scale=0.60]{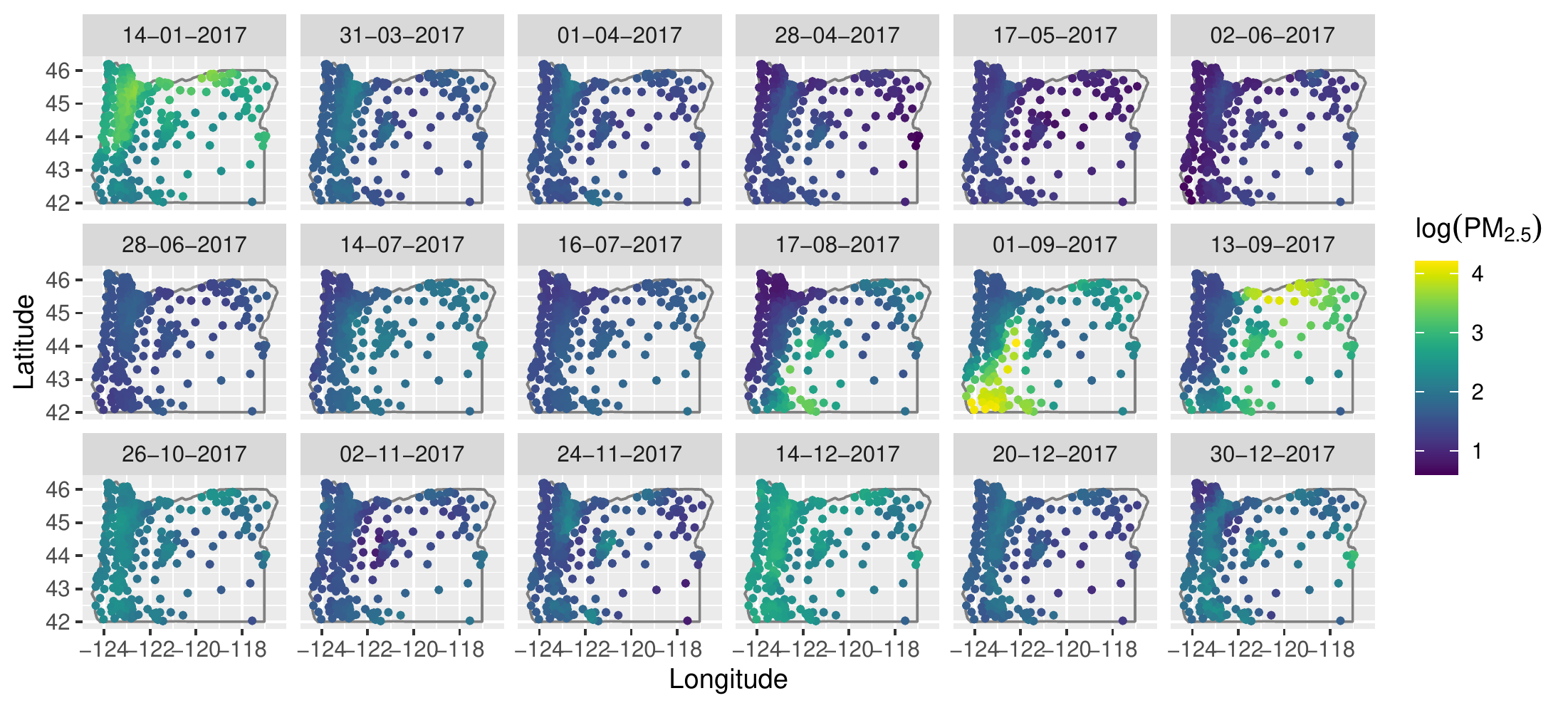}
\caption{Observed log(PM$_{2.5}$) data over the state of Oregon, US, for 18 randomly selected days in the year 2017.}
\label{fig1}
\end{figure}

The PM$_{2.5}$ data in consideration is sourced from the Environmental Protection Agency (EPA) which provides the daily average of PM$_{2.5}$, spatially spanning the United States. The PM$_{2.5}$ data from the EPA is generated by integrating the monitoring data from National Air Monitoring Stations/State and Local Air Monitoring Stations (NAMS/SLAMS) with 12 km gridded output from the Community Multiscale Air Quality (CMAQ) (\url{https://www.epa.gov/cmaq}) modeling system. While the spatial coverage of the raw dataset extends across the entire United States, we focus our analysis only on the state of Oregon and consider the PM$_{2.5}$ data for the year 2017. This specific choice of the study region is driven by the fact that Oregon is one of the states in US which suffer from wildfires, and the study of its PM$_{2.5}$ concentration is of particular interest. The total volume of Oregon's space-time PM$_{2.5}$ data equals 293,825 observations in the form of daily time series at 805 two dimensional spatial locations.  In terms of probability distributions, the PM$_{2.5}$ data exhibit positively skewed distribution 
 and the corresponding log transformation is nearly Gaussian.  
 Therefore, we choose to analyze log(PM$_{2.5}$) instead of PM$_{2.5}$ since the former closely satisfies the Gaussian process assumption. Figure~\ref{fig1} visualizes the spatial fields of PM$_{2.5}$ at Oregon on a logarithmic scale for 18 randomly selected days from the year 2017.  Clearly, the 805 observed spatial locations shown in Figure \ref{fig1} are not uniformly distributed across the state of Oregon as the observed locations are mainly concentrated in west, north-west and south-west, which are the regions with high population density, whereas the eastern region, for the most part, is unobserved.   
Therefore, the main objective of our analysis is to satisfactorily model the spatio-temporal dependence of the considered data so as to develop an accurate predictive model which can continuously predict PM$_{2.5}$ in the unobserved part of the study region. 

The accuracy of any predictive model needs to be evaluated by means of cross-validation, and accordingly, we divide the data into training and validation data. We consider two sets of validation data where the first set consists of data at all 805 locations from day 356 to day 365, and the second set consists of data at 483 randomly selected locations (60\% of the total locations) from day 1 to day 355. The first and second set of validation data are used to gauge the forecasting and interpolation accuracy of predictive models, respectively, and therefore are referred to as ``forecasting validation set'' and ``interpolation validation set'', respectively. The observed data at the remaining 322 locations from day 1 to day 355 constitute the training data.

\begin{figure}[h]
\centering     
\includegraphics[scale=0.60]{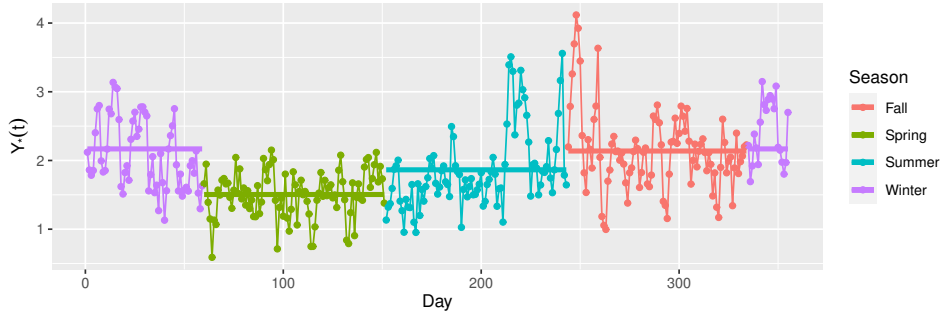}
\caption{Spatially averaged log(PM$_{2.5}$) for Day 1 to Day 355 of the year 2017. The four different colors represent four seasons namely: fall, spring, summer and winter. The horizontal solid line of the four different colors represent the four seasonal means.}
\label{fig3}
\end{figure}

Let $Y(\s,t), \: (\s,t)\in\mathbb{R}^2\times\mathbb{R},$ denote the log(PM$_{2.5}$) observed at spatial location $\s=(\text{Longitude, Latitude})$ and the time $t= (\text{day of the year 2017}-1)/(365-1)$, 
 and $\mathcal{S}_T$ denote the set of the 322 training locations. Note that the time $t$ is scaled such that $t=0$ for the first day and $t=1$ for the last day of the year 2017. We aim to model $Y(\s,t)$ as a spatio-temporal Gaussian process such that $\mathbb{E}\{Y(\s,t)\}=\mu(\s,t), (\s,t)\in\mathbb{R}^2\times\mathbb{R}$ and $\text{Cov}\{Y(\s_1,t_1),Y(\s_2,t_2)\}=K(\s_1,\s_2,t_1,t_2)$. We model the mean function $\mu(\s,t)$ as a linear function of multiple terms which are functions of $\s$ and $t$ to capture the spatio-temporal trends and seasonality. To decide the terms in the linear function, we first investigate any potential presence of seasonality through an exploratory time series plot of $Y_*(t)$ vs. $364t+1$ (Day) in Figure \ref{fig3}, where $Y_*(t)=\sum_{\s \in \mathcal{S}_T}Y(\s,t)/322,$ represents the spatial average of log(PM$_{2.5}$) over training locations on a given day of the year. The time series plot is segmented into four seasons, namely fall (September, October, November), spring (March, April, May), summer (June, July, August) and winter (December, January, February); with their respective seasonal averages shown as well. The plot indicates the potential seasonality effect as $Y_*(t)$ values are generally higher in fall and winter, and lower in spring and summer; which in turn suggests to consider harmonic terms of $t$ while modeling $\mu(\s,t)$. Additionally, to incorporate spatio-temporal trend in $\mu(\s,t)$, we also include the direct and interaction terms of $\s$ and $t$. Finally, we assume and fit  the following linear model for $\mu(\s,t)$ on the training data: 
\begin{multline*}
\mu(\s,t)=\beta_0+\beta_1\sin\Big(\frac{2\pi t}{0.5}\Big)+\beta_2\cos\Big(\frac{2\pi t}{0.5}\Big)+\beta_3\sin\Big(\frac{4\pi t}{0.25}\Big)+\beta_4\cos\Big(\frac{4\pi t}{0.25}\Big)+\beta_5t+\\\boldsymbol{\beta_6}\s^\text{T}+\boldsymbol{\beta_7}\s^\text{T}t+\boldsymbol{\beta_8}\s^\text{T}t^2+\boldsymbol{\beta_9}\s^\text{T}t^3+\boldsymbol{\beta_{10}}\s^\text{T}t^4,
\end{multline*} where $\beta_0$ is the intercept, $\{\beta_1,\beta_2,\beta_3, \beta_4\}$ are the seasonality coefficients, $\beta_5$ is the temporal trend, $\boldsymbol{\beta_6}$ is the vector of spatial trend coefficients and $\{\boldsymbol{\beta_7},\boldsymbol{\beta_8},\boldsymbol{\beta_9},\boldsymbol{\beta_{10}}\}$ are the space-time interaction coefficients.

\begin{figure}[h]
\centering     
\subfigure[]{\label{fig4a}\includegraphics[width=60mm]{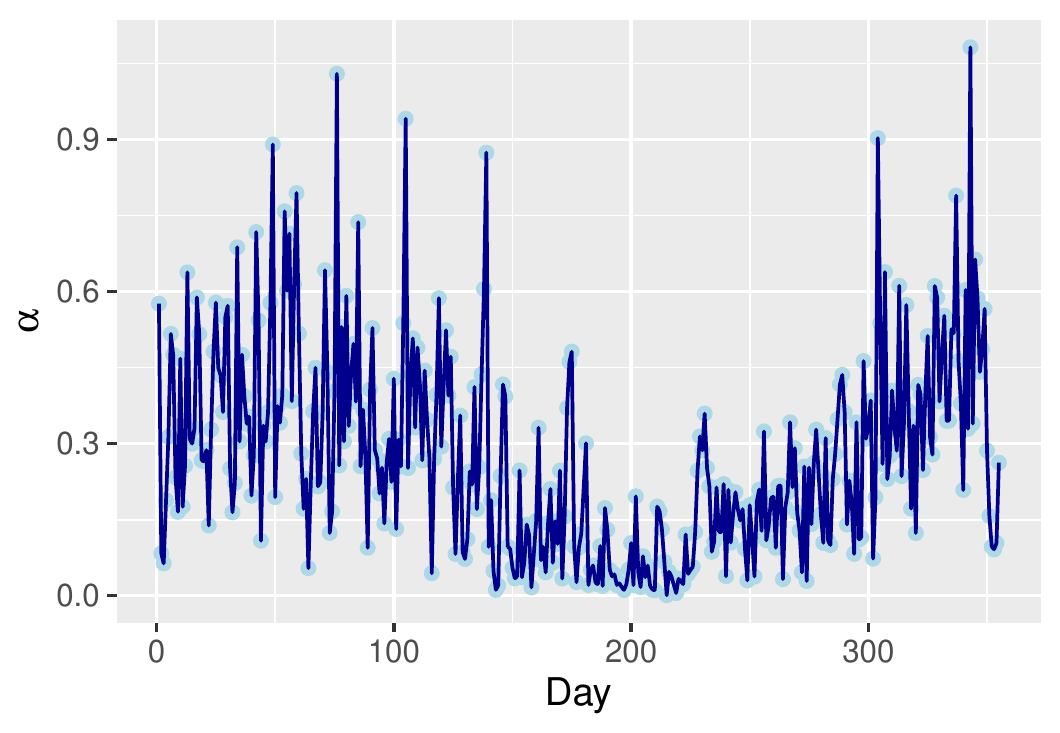}}
\subfigure[]{\label{fig4b}\includegraphics[width=60mm]{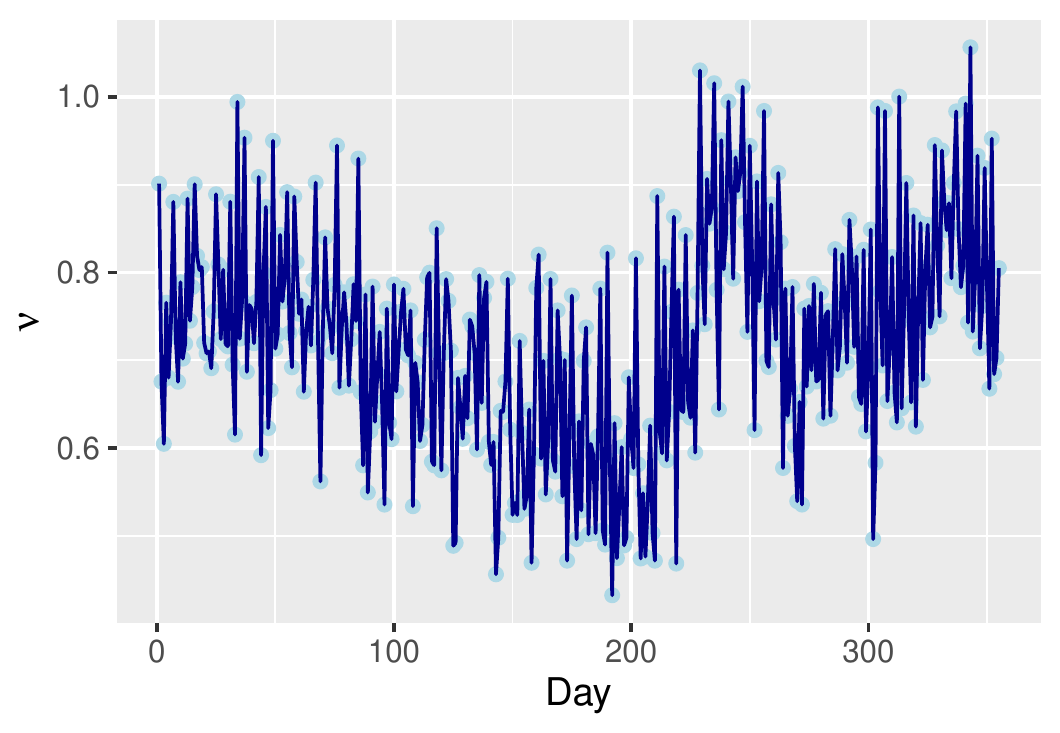}}
\caption{Maximum likelihood estimates of the (a) spatial scale parameter $\alpha$, and (b) smoothness parameter $\nu$, of the Mat\'ern correlation function, fitted independently over spatial field of residuals for each day of the training data. Note that the estimates for $\alpha$ for day 180 and 218 are 2.32 and 1.46, respectively, and those estimates are skipped in the above plot to highlight the main trend.  }
\label{fig4}
\end{figure}

We then proceed to detrend the process $Y(\s,t)$ with the fitted $\hat{\mu}(\s,t)$ and obtain the residual process $\epsilon(\s,t)=Y(\s,t)-\hat{\mu}(\s,t)$, which is further investigated to explore the properties of $K(\s_1,\s_2,t_1,t_2)$. For simplicity, we assume that the process is stationary in space, and therefore, the covariance function $K(\s_1,\s_2,t_1,t_2)=\text{C}(\s_2-\s_1,t_1,t_2),$ depends only on spatial lag $\s_2-\s_1$ and time-points $t_1,t_2$. Furthermore, we also assume that the purely spatial covariance function for any arbitrary time $t_i$: $\text{C}(\s_2-\s_1,t_i,t_i)$ is of the Mat\'ern class. Next, we want to explore the time-varying properties of  $\text{C}(\s_2-\s_1,t_i,t_i)$; and therefore, we independently fit a purely spatial Mat\'ern covariance function $\sigma^2\text{M}(\textbf{h}\mid \alpha, \nu)$, for day 1 to day 355, on the corresponding training sample of $\epsilon(\s,t)$, using the maximum likelihood estimation (MLE) method. Figure \ref{fig4} shows the daywise estimates of spatial scale parameter $\alpha$ (see Figure \ref{fig4a}) and smoothness parameter $\nu$ (see Figure \ref{fig4b}) from the fitted spatial Mat\'ern covariance function. There appears to be an obvious temporal evolution in both the parameters as both $\alpha$ and $\nu$ exhibit an increasing trend in the second half of the year and a predominantly decreasing trend in the first half of the year.  
Therefore, this time-varying spatial dependence must be taken into account while specifying the spatio-temporal covariance function $K(\s_1,\s_2,t_1,t_2)$. Failing to do so can cause misspecification of the process and might lead to sub-optimal inference and prediction. This motivates the construction of our time-varying class of spatio-temporal covariance functions, details of which are given in the section \ref{method}. The data analysis shown here is further continued in Section \ref{dataanalysis}. 

\section{Covariance Model Construction and Estimation}\label{method}
In this section, we introduce our proposed class of time-varying spatio-temporal covariance functions with discussion on its properties and validity conditions (see Section \ref{tvarmethod}). Besides, we also discuss the random composite likelihood-based estimation method (see Section \ref{rclmethod}) that we implement in our simulation study and data application.
\subsection{Time-varying spatio-temporal covariance models}\label{tvarmethod}
We consider the Gneiting-Mat\'ern class of nonseparable stationary spatio-temporal covariance functions \eqref{eq2}, and provide its time-varying generalization in the following theorem:

\begin{theorem}\label{theorem1}
Let $\alpha_s(t)>0, t\in \mathbb{R}$ and $\nu_s(t)>0, t\in\mathbb{R},$ be any positive real valued functions, then the following time-varying spatio-temporal covariance function in \eqref{eq3}:
\begin{eqnarray}\label{eq3}
     \text{Cov}\{X(\textbf{s},t_i),X(\textbf{s}+\textbf{h},t_j)\}=\text{C}(\h,t_i,t_j)=\sigma^2\frac{\Gamma\{\frac{\nu_s(t_i)+\nu_s(t_j)}{2}\}}{\sqrt{\Gamma\{\nu_s(t_i)\}\Gamma\{\nu_s(t_j)\}}}\times\\\nonumber\frac{1}{\{\alpha_s^2(t_i)\}^{d/4}\{\alpha_s^2(t_j)\}^{d/4}\{\frac{\psi(|t_i-t_j|^2)}{\overline{\alpha}_s^2}+\frac{1/\alpha^2_s(t_i)+1/\alpha^2_s(t_j)}{2}-\frac{\psi(0)}{\overline{\alpha}_s^2}\}^{d/2}}\times\\\nonumber\text{M}[\textbf{h}\mid\frac{1}{\{\frac{\psi(|t_i-t_j|^2)}{\overline{\alpha}_s^2}+\frac{1/\alpha^2_s(t_i)+1/\alpha^2_s(t_j)}{2}-\frac{\psi(0)}{\overline{\alpha}_s^2}\}^{1/2}},\frac{\nu_s(t_i)+\nu_s(t_j)}{2}],
    \end{eqnarray}
is valid for any Bernstein function $\psi(w)>0,\; w\geq 0,$ and $\overline{\alpha}_s>0$. 
\end{theorem}

The proof of Theorem \ref{theorem1} materializes by reckoning spatio-temporal processes as multivariate spatial processes, details of which are deferred to the Supplementary Material. 

While the spatio-temporal covariance function in \eqref{eq3} is valid for any positive value of the parameter $\overline{\alpha}_s$, we now onwards choose to constrain it as $\overline{\alpha}_s=\sum_{t_i\in\text{T}}\alpha_s(t_i)/\text{T}$, where T represents the set of all the training time-points. This constraint is beneficial for two reasons: (i) it renders a simpler model with one less parameter to be estimated, and (ii) include \eqref{eq2} as a special case when $\alpha_s(t_i)$ and $\nu_s(t_i)$ are constant over time. Specifically, with any standardized Bernstein function $\psi(w),w\geq0, \psi(0)=1, $ let $\alpha_s(t)=\alpha>0,\: \nu_s(t)=\nu>0,\: t\in \mathbb{R},$ in \eqref{eq3}, then $\overline{\alpha}_s=\alpha$ and \eqref{eq3} reduces to:  
\begin{eqnarray}
\nonumber\text{C}(\h,t_i,t_j)=\frac{\sigma^2}{\{{\psi(|t_i-t_j|^2)}\}^{d/2}}\Bigg(\frac{\alpha\|\h\|}{\psi(|t_i-t_j|^2)^{1/2}}\Bigg)^\nu K_{\nu}\Bigg(\frac{\alpha\|\h\|}{\psi(|t_i-t_j|^2)^{1/2}}\Bigg),\end{eqnarray}
which is a Gneiting-Mat\'ern class \eqref{eq2}, and on that account, \eqref{eq3} is a time-varying generalization of \eqref{eq2}. The time-varying properties of the spatio-temporal covariance model in \eqref{eq3} become intelligible in its purely spatial and purely temporal restrictions. In particular, let $t_i=t_j=t'$ in \eqref{eq3} to evaluate the purely spatial covariance at any arbitrary time point $t'$, we get: $\text{C}(\h,t',t')=\sigma^2\text{M}\{\textbf{h}\mid\alpha_s(t'),\nu_s(t')\},$ which is a spatial Mat\'ern covariance function with spatial scale $\alpha_s(t')$ and smoothness $\nu_s(t')$. Accordingly, the functions $\alpha_s(t)$ and $\nu_s(t)$ represent the spatial scale and smoothness of the purely spatial Mat\'ern covariance function in \eqref{eq3} at any given time point $t$, thus allowing the temporal evolution of spatial dependence. As an immediate consequence of functions $\alpha_s(t)$ and $\nu_s(t)$, the following purely temporal restriction of \eqref{eq3} becomes nonstationary in time: 
\begin{equation}\label{eq4}
\text{C}(\boldsymbol{0},t_i,t_j)=\frac{\sigma^2\Gamma\{\frac{\nu_s(t_i)+\nu_s(t_j)}{2}\}\{\alpha_s^2(t_i)\}^{-d/4}\{\alpha_s^2(t_j)\}^{-d/4}}{\sqrt{\Gamma\{\nu_s(t_i)\}\Gamma\{\nu_s(t_j)\}}\{\frac{\psi(|t_i-t_j|^2)}{\overline{\alpha}_s^2}+\frac{1/\alpha^2_s(t_i)+1/\alpha^2_s(t_j)}{2}-\frac{\psi(0)}{\overline{\alpha}_s^2}\}^{d/2}}.
\end{equation}

While the nonstationary behavior of the purely temporal covariance in \eqref{eq4} is entirely controlled by nontrivial interactions of functions $\alpha_s(t)$ and $\nu_s(t)$, the individual interpretation of those functions in the context of temporal nonstationarity becomes clear when we vary them singly in \eqref{eq4}. Firstly, let us fix: (i) $\nu_s(t)=\nu_f,\; t\in \mathbb{R},$  (ii) $\alpha_s(t_r)=\alpha_r;\: \alpha_s(t_j)=\alpha_f,\: t_j\neq t_r \in \mathbb{R},$ for any arbitrary reference time point $t_r$,  and (iii) $\overline{\alpha}_s=\alpha_f$ in \eqref{eq4}, then the temporal covariance of the reference time-point $t_r$ with any other time point $t_j$ is given as:
\begin{equation}\label{eq5}
\text{C}(\boldsymbol{0},t_r,t_j)=\frac{\sigma^2(\frac{\alpha_f^2}{\alpha_r^2})^{d/4}}{\{\psi(|t_r-t_j|^2)-\psi(0)+\frac{1}{2}(1+\frac{\alpha_f^2}{\alpha_r^2})\}^{d/2}},\ \text{for all} \  t_j\in\mathbb{
R}.\end{equation}
Note that it is reasonable here to fix $\overline{\alpha}_s=\alpha_f$ even under the aforementioned choice of constraint for $\overline{\alpha}_s$, as $\overline{\alpha}_s=\sum_{t_i\in\text{T}}\alpha_s(t_i)/\text{T}\approx\alpha_f,$ provided that T includes a large number of training points and $\alpha_r$ is not extremely different from $\alpha_f$.  
The function in \eqref{eq5} expresses covariance of a reference time point $t_r$ with any other time point $t_j\in\mathbb{R}$ as a function of $\mathbb{L}_1$ distance between them, i.e., $|t_r-t_j|$ , and the term $\alpha_f/\alpha_r$  counter-balances the scale of the covariance and  the rate of covariance decay with increasing distance $|t_r-t_j|$. For instance, if $\alpha_r<\alpha_f$, then for non-zero temporal lags, the scale of covariance is increased and the rate of covariance decay is decreased through the term $\alpha_f/\alpha_r$ in the numerator and denominator of \eqref{eq5}, respectively. Therefore, the function $\alpha_s(t),$ not only denotes the spatial scale of purely spatial Mat\'ern covariance at time $t$, but also governs the scaling and rate of temporal covariance decay away from the time point $t$. Next, we fix (i) $\alpha_s(t)=\alpha_f, \ t\in \mathbb{R}$, (ii) $\overline{\alpha}_s=\alpha_f$, and (iii) $\nu(t_r)=\nu_r; \nu(t_j)=\nu_f, t_j\neq t_r\in\mathbb{R}$ in \eqref{eq4}, we get: 
\begin{equation}\label{eq6}
\text{C}(\boldsymbol{0},t_r,t_j)=\frac{\sigma^2\Gamma(\frac{\nu_r+\nu_f}{2})}{\sqrt{\Gamma(\nu_r)\Gamma(\nu_f)}\{\psi(|t_r-t_j|^2)-\psi(0)+1\}^{d/2}},\ \text{for all} \  t_j\in\mathbb{
R,}\end{equation}
where the terms $\nu_r$ and $\nu_f$ control the scale of the temporal covariance such that the covariance is scaled down if $\nu_r\neq\nu_f$ and the magnitude of this downscaling is directly proportional to the difference between $\nu_f$ and $\nu_r$. Hence, the function $\nu_s(t)$ also plays a twofold role where on one hand it controls the smoothness of the purely spatial Mat\'ern covariance at time $t$, on the other hand, it regulates the scaling of temporal covariance at non-zero temporal lags.

\begin{figure}
\centering     
\subfigure[]{\label{fig5a}\includegraphics[width=60mm]{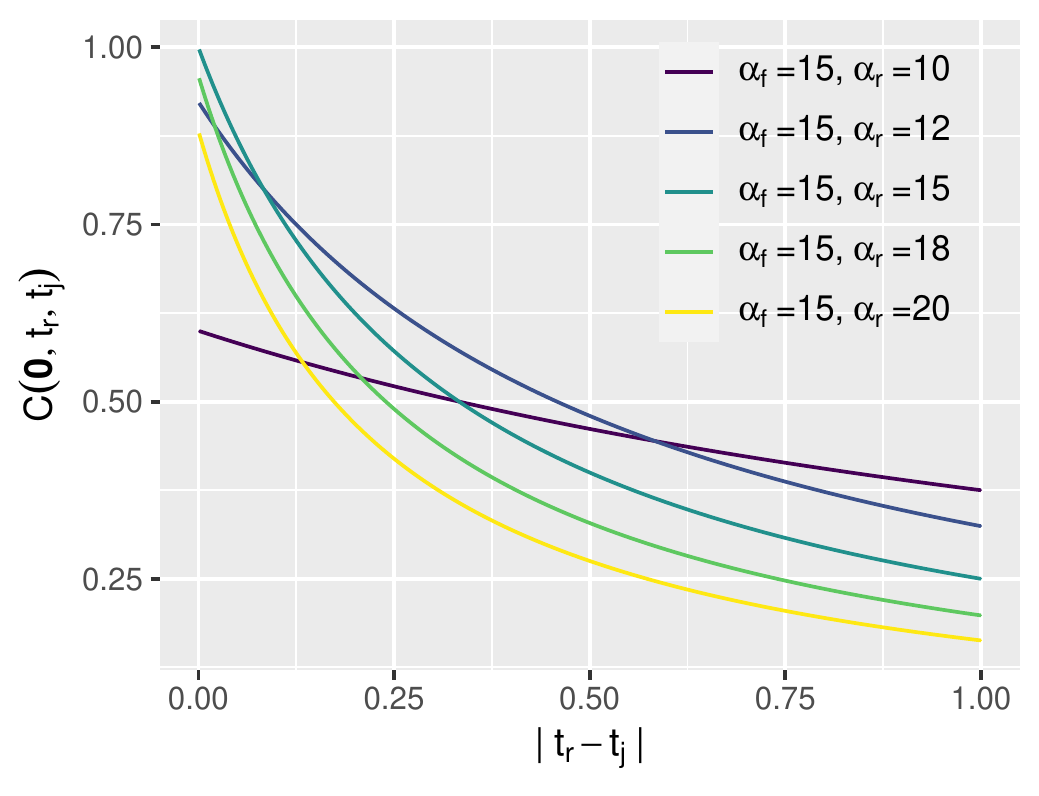}}
\subfigure[]{\label{fig5b}\includegraphics[width=60mm]{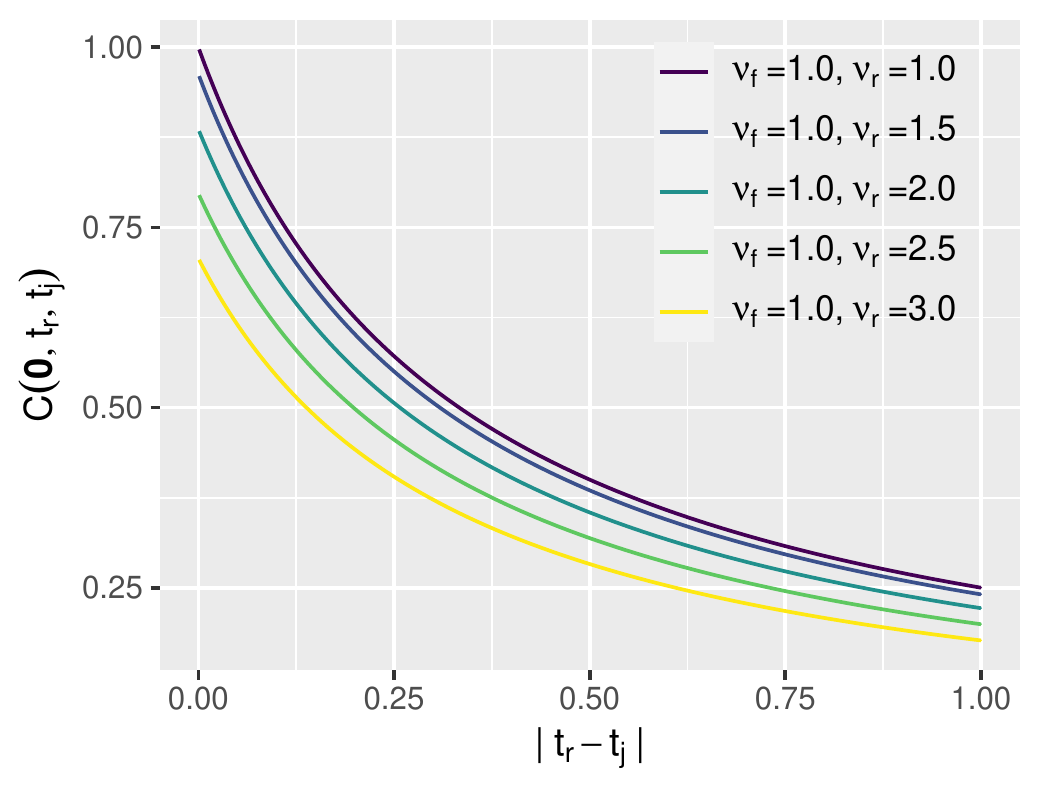}}
\caption{(a) The purely temporal covariance function in \eqref{eq5} as a function of $|t_r-t_j|$ for different choices of $\alpha_f$ and $\alpha_r.$ (b) The purely temporal covariance function in \eqref{eq6} as a function of $|t_r-t_j|$ for different choices of $\nu_f$ and $\nu_r.$ For both (a) and (b), we have fixed $\sigma=1$ and $\psi(w)=(3w^{0.5}+1).$  }
\label{fig5}
\end{figure}

These effect of $\alpha_s(t)$ and $\nu_s(t)$ on the purely temporal covariance function is also illustrated with examples in Figure \ref{fig5}. For $\sigma=1$ and $\psi(w)=(3w^{0.5}+1)$, Figure \ref{fig5a} and Figure \ref{fig5b} show the temporal covariance function in \eqref{eq5} for different combinations of $(\alpha_f,\alpha_r)$ and the temporal covariance function in \eqref{eq6}, for different combinations of $(\nu_f,\nu_r)$, respectively. In particular,  the illustrated combinations are $(\alpha_f,\alpha_r)\in\{(15,10),(15,12),(15,15),(15,18),(15,20)\}$ and $(\nu_f,\nu_r)\in\{(1,1), (1,1.5), (1,2), (1,2.5),$ $(1,3$\\$)\}$, where we consider $(\alpha_f,\alpha_r)=(15,15)$ and $(\nu_f,\nu_r)=(1,1)$ as the base cases for studying their effects. As shown in Figure \ref{fig5a}, for the cases when $\alpha_f>\alpha_r$, i.e., $(\alpha_f,\alpha_r)\in\{(15,10),(15,12)\}$, the rate of covariance decay is decreased and the scaling is increased compared to the base case, whereas for the cases $\alpha_f<\alpha_r$, i.e., $(\alpha_f,\alpha_r)\in\{(15,18),(15,20)\}$, the rate of covariance decay is increased and the scaling is decreased. Moreover, the effect is stronger when the difference between $\alpha_f$ and $\alpha_r$ is higher. Similarly, relative to the base case, the scale of the covariance is clearly decreased in Figure \ref{fig5b} for $(\nu_f,\nu_r)\in\{ (1,1.5), (1,2), (1,2.5),$ $(1,3)\}$ and the decrease is the highest when $\nu_r$ is the farthest from $\nu_f$, i.e., $(\nu_f,\nu_r)=(1,3)$.

The proper specification of functional forms for $\alpha_s(t)$ and $\nu_s(t)$ in \eqref{eq3}, which can flexibly capture the time-varying dependence of any considered spatio-temporal data, is consequential to achieve an advantageous modeling and inference. Thus, the definition of the functions $\alpha_s(t)$ and $\nu_s(t)$ should ideally be based on empirical evidence from some exploratory data analysis. One such alternative is to follow the analysis of Section \ref{eda} and utilize the time series plots analogous to Figure~\ref{fig4} for defining the functions $\alpha_s(t)$ and $\nu_s(t)$. Specifically, Figures \ref{fig4a} and \ref{fig4b} can guide the choice for functional forms of $\alpha_s(t)$ and $\nu_s(t)$, respectively, as those time series are indeed an empirical counterpart of the corresponding functions 
$\alpha_s(t)$ and $\nu_s(t)$. While there can be innumerable possible constructions of positive real-valued functions which are eligible choices for $\alpha_s(t)$ and $\nu_s(t)$, we consider the following definition of $\alpha_s(t)$ and $\nu_s(t)$ for the time-varying model estimation in our work: $\alpha_s(t)=\exp(p^\alpha_{n.\alpha}(t)), n.\alpha\geq0,$ and $\nu_s(t)=\exp(p^\nu_{n.\nu}(t)), n.\nu\geq0$, where $p^\alpha_k(t)$ and $p^\nu_k(t)$, both are $k-$order polynomial of $t$. 

For the choice of $\psi(w)>0, w\geq0$ in \eqref{eq3}, there are several options available from the list of Bernstein functions given in \cite{van2012potential} and \cite{gneiting2002}. In this work, we only consider the following choice of Bernstein function: $\psi(w)=(aw^\gamma+1)^\beta,\ a>0, \ 0<\gamma\leq 1, \ 0\leq \beta \leq 1$. Moreover, to impart additional flexibility in the temporal part similar to Example 2 of \cite{gneiting2002}, we multiply \eqref{eq3} with a purely temporal covariance function: $(a|t_i-t_j|^{2\gamma}+1)^{-\delta}, \delta\geq 0,$ where $a$ and $\gamma$ are the parameters common to the chosen function $\psi(w)$. Consequently, the time-varying spatio-temporal model in \eqref{eq3} reduces to:
\begin{eqnarray}\label{eq7}
     \text{C}(\h,t_i,t_j)=\sigma^2\frac{\Gamma\{\frac{\nu_s(t_i)+\nu_s(t_j)\}}{2}\}}{\sqrt{\Gamma\{\nu_s(t_i)\}\Gamma\{\nu_s(t_j)\}}}\times\\\nonumber\frac{1}{\{\alpha_s^2(t_i)\}^{d/4}\{\alpha_s^2(t_j)\}^{d/4}\{\frac{(a|t_i-t_j|^{2\gamma}+1)^\beta}{\overline{\alpha}_s^2}+\frac{1/\alpha^2_s(t_i)+1/\alpha^2_s(t_j)}{2}-\frac{1}{\overline{\alpha}_s^2}\}^{d/2}{(a|t_i-t_j|^{2\gamma}+1)^\delta}}\times\\\nonumber\text{M}[\textbf{h}\mid \frac{1}{\{\frac{(a|t_i-t_j|^{2\gamma}+1)^\beta}{\overline{\alpha}_s^2}+\frac{1/\alpha^2_s(t_i)+1/\alpha^2_s(t_j)}{2}-\frac{1}{\overline{\alpha}_s^2}\}^{1/2}},\frac{\nu_s(t_i)+\nu_s(t_j)\}}{2}],
    \end{eqnarray}
where $\sigma>0,\ \delta\geq0,\ 0\leq\beta\leq 1,\ a>0, \ 0<\gamma\leq 1,\ \alpha_s(t)>0, \nu_s(t)>0, t\in\mathbb{R},$ and $\overline{\alpha}_s=\sum_{t_i\in\text{T}}\alpha_s(t_i)$. Now, if we set $\alpha_s(t)=\alpha, t\in \mathbb{R}$ and $\nu_s(t)=\nu, t\in \mathbb{R}$, then \eqref{eq7} reduces to the following Gneiting-Mat\'ern class:
\begin{eqnarray}\label{eq8}
     \text{C}(\h,t_i,t_j)=\sigma^2\frac{1}{(a|t_i-t_j|^{2\gamma}+1)^{\beta d/2+\delta}}\text{M}\{\textbf{h}\mid \frac{\alpha}{(a|t_i-t_j|^{2\gamma}+1)^{\beta /2}},\nu\},
    \end{eqnarray}
   where $\beta$ represents the parameter to control the degree of nonseparability such that $\beta=1$ corresponds to fully nonseparable model and $\beta=0$ leads to the following separable model:
\begin{eqnarray}\label{eq9}
     \text{C}(\h,t_i,t_j)=\sigma^2\frac{1}{(a|t_i-t_j|^{2\gamma}+1)^{\delta}}\text{M}(\textbf{h}\mid \alpha,\nu).
    \end{eqnarray}

The three spatio-temporal covariance models in \eqref{eq7}, \eqref{eq8} and \eqref{eq9} are considered as the candidate models in the simulation study and data application presented in Section \ref{simulation} and Section \ref{dataanalysis}, respectively. The model in \eqref{eq7} can be made arbitrarily flexible through parametric functions for $\alpha_s(t)$ and $\nu_s(t)$, however, the estimation of model parameters through Gaussian MLE also then becomes increasingly challenging. In principle, large volume of data is preferred to fit a highly parameterized complex model such as \eqref{eq7} to avoid over-fitting, but at the same time the Gaussian MLE becomes time-prohibitive and computationally infeasible with a high volume of spatio-temporal data. The main issue lies in storing and performing the Cholesky factorization of the large spatio-temporal covariance matrix. To overcome this, we implement a composite likelihood-based estimation procedure which is described in Section \ref{rclmethod}.

\subsection{Random composite likelihood estimation}\label{rclmethod}

Let $X(\s,t), \textbf{s}\in\mathbb{R}^d, t \in \mathbb{R}$ be a zero mean Gaussian spatio-temporal process, and $\boldsymbol{X}_{{\mathcal{S}},{\mathcal{T}}}$ denote the vector of the process $X,$ observed at the set of locations $\mathcal{S}=\{\s_1,\ldots,\s_{ns}\}\subset\mathbb{R}^d$, $ns\geq1,$ and the set of time-points $\mathcal{T}=\{t_1,\ldots,t_{nt}\}\subset\mathbb{R}$, $nt\geq 1,$ i.e., $\boldsymbol{X}_{{\mathcal{S}},{\mathcal{T}}}=\{X(\s,t); \s \in \mathcal{S}, t \in \mathcal{T}\}.$ The total number of data points is denoted as $N=ns\cdot nt$. The log-likelihood function for $\boldsymbol{X}_{{\mathcal{S}},{\mathcal{T}}}$ is given as: $\ell(\boldsymbol{\theta}\mid \boldsymbol{X}_{{\mathcal{S}},{\mathcal{T}}})=-\{\log\det\Sigma(\boldsymbol{\theta})+\boldsymbol{X}_{{\mathcal{S}},{\mathcal{T}}}^\text{T}\Sigma(\boldsymbol{\theta})^{-1}\boldsymbol{X}_{{\mathcal{S}},{\mathcal{T}}}+N.\log 2\pi\}/2,$ where $\Sigma(\boldsymbol{\theta})$ is the $N\times N$ covariance matrix for $\boldsymbol{X}_{{\mathcal{S}},{\mathcal{T}}}$, defined through a spatio-temporal covariance function which depends on the set of parameters $\boldsymbol{\theta}$. The maximum likelihood estimation of $\boldsymbol{\theta}$ requires computing: $\hat{\boldsymbol{\theta}}_{ML}=\argmaxE_{\boldsymbol{\theta}}\ell(\boldsymbol{\theta}\mid \boldsymbol{X}_{{\mathcal{S}}{\mathcal{T}}})$, generally done  through numerical optimization routines which involve iterative evaluation of $\ell(\boldsymbol{\theta}\mid \boldsymbol{X}_{{\mathcal{S}}{\mathcal{T}}})$. The optimization becomes computationally challenging in case both or either of $ns$ and $nt$ are large, as  $\Sigma(\boldsymbol{\theta})$ then becomes a large covariance matrix and the iterative evaluation of  $\ell(\boldsymbol{\theta}\mid \boldsymbol{X}_{{\mathcal{S}},{\mathcal{T}}})$ becomes time-prohibitive. Additionally, storing an extremely large sized covariance matrix $\Sigma(\boldsymbol{\theta})$ can exhaust the available memory of the machine, thus making the optimization impracticable. A widely used approximate solution to curtail this computational issue is to adopt composite likelihood methods \citep{vecchia1988,steincomposite,varin2011overview,Eidsvik}, in which the optimization is carried out over the product of component likelihoods. 

In this work, we too implement the estimation by the means of composite likelihood where the collection of component likelihoods is chosen randomly. Specifically, we randomly create equisized subsets $\mathcal{S}_{ij}\subset\mathcal{S},\ i=1,\ldots, R_s,\ j=1,\ldots,M_S,$ and $\mathcal{T}_{ij}\subset\mathcal{T},\ i=1,\ldots, R_t,\ j=1,\ldots,M_t,$  such that for each $i=1,\ldots,R_s$: $\cup_{j=1}^{M_s}\mathcal{S}_{ij}=\mathcal{S}, \ \mathcal{S}_{ik}\cap\mathcal{S}_{il}=\phi, \forall \ k\neq l$, and for each $i=1,\ldots,R_t$: $\cup_{j=1}^{M_t}\mathcal{T}_{ij}=\mathcal{T}, \ \mathcal{T}_{ik}\cap\mathcal{T}_{il}=\phi, \forall \ k\neq l$. Here, $M_s$ and $M_t$ govern the size of subsets of $\mathcal{S}$ and $\mathcal {T}$, respectively, whereas $R_s$ and $R_t$ denote the number of randomly created mutually exclusive and exhaustive partitions of $\mathcal{S}$ and $\mathcal{T}$, respectively. Based on those subsets, we define the following random composite log-likelihood (RCL) function:
\begin{equation}\label{eq10}
     \ell_{RC}(\boldsymbol{\theta}\mid \boldsymbol{X}_{\mathcal{S},\mathcal{T}} )=\frac{\sum_{i=1}^{R_s}\sum_{j=1}^{M_s}\ell(\boldsymbol{\theta}\mid \boldsymbol{X}_{{\mathcal{S}_{ij}},{\mathcal{T}}})}{2}+\frac{\sum_{i=1}^{R_t}\sum_{j=1}^{M_t}\ell(\boldsymbol{\theta}\mid \boldsymbol{X}_{{\mathcal{S}},{\mathcal{T}_{ij}}})}{2},
\end{equation} and the RCL estimate of $\boldsymbol{\theta}$ is then obtained as $\hat{\boldsymbol{\theta}}_{RCL}=\argmaxE_{\boldsymbol{\theta}}\ell_{RC}(\boldsymbol{\theta}\mid \boldsymbol{X}_{{\mathcal{S}}{\mathcal{T}}})$. For large spatio-temporal datasets, computation and optimization of $\ell_{RC}(\boldsymbol{\theta}\mid \boldsymbol{X}_{{\mathcal{S}}{\mathcal{T}}})$ is relatively more feasible than that of $\ell(\boldsymbol{\theta}\mid \boldsymbol{X}_{{\mathcal{S}}{\mathcal{T}}})$ as the former includes smaller-sized covariance matrices because the component log-likelihoods are based only on the subset of the data. Additionally, $\ell_{RC}(\boldsymbol{\theta}\mid \boldsymbol{X}_{{\mathcal{S}}{\mathcal{T}}})$ can also easily utilize the parallel architecture of modern machines to simultaneously compute the component log-likelihoods, which would lead to further computational speed up.  The functions  $\ell_{RC}(\boldsymbol{\theta}\mid \boldsymbol{X}_{{\mathcal{S}}{\mathcal{T}}})$ and $\ell(\boldsymbol{\theta}\mid \boldsymbol{X}_{{\mathcal{S}}{\mathcal{T}}})$ become increasingly similar for smaller values $M_s$ and $M_t$, therefore, smaller $M_s$ and $M_t$ leads to more accurate but slower estimation. Note that if $M_s=M_t=1$, then $\ell_{RC}(\boldsymbol{\theta}\mid \boldsymbol{X}_{{\mathcal{S}}{\mathcal{T}}})=\ell(\boldsymbol{\theta}\mid \boldsymbol{X}_{{\mathcal{S}}{\mathcal{T}}})$ as $M_s=1\implies R_s=1$ and $M_t=1\implies R_t=1$. Therefore, the values of $M_s,M_t,R_s$ and $R_t$ should be chosen by considering the trade-off between accuracy and speed. 

We provide an exposition on the properties of $\ell_{RC}(\boldsymbol{\theta}\mid \boldsymbol{X}_{{\mathcal{S}},{\mathcal{T}}})$  in the Supplementary Material, wherein, we prove that the random composite likelihood score function is always an unbiased estimating function for $\boldsymbol{\theta}$, i.e., $\mathbb{E}\{\frac{\partial \ell_{RC}(\boldsymbol{\theta}\mid \boldsymbol{X}_{{\mathcal{S}},{\mathcal{T}}})}{\partial \theta_r}\}=0$. In addition, we have also included the evaluation for the Hessian of $\ell_{RC}(\boldsymbol{\theta}\mid \boldsymbol{X}_{{\mathcal{S}},{\mathcal{T}}})$ and the variance of $\hat{\boldsymbol{\theta}}_{RCL}$.

\section{Simulation Study}\label{simulation}
In this section, we conduct a simulation study to empirically evaluate the advantage of using the proposed time-varying class of spatio-temporal covariance models against the commonly used Gneiting-Mat\'ern class and the separable class of spatio-temporal covariance models. Moreover, we enact and assess the RCL estimation for the three classes of models. For this simulation study, we particularly consider the three nested models  \eqref{eq7}, \eqref{eq8} and  \eqref{eq9}, which now onwards, are referred as ``Tvar.M'', ``Gneit.M'' and ``Sep.M'', respectively, for brevity. These models are compared on the basis of interpolation and forecasting performance under four different cases of simulated spatio-temporal Gaussian processes.

\begin{figure}[h]
\centering     
\includegraphics[scale=0.55]{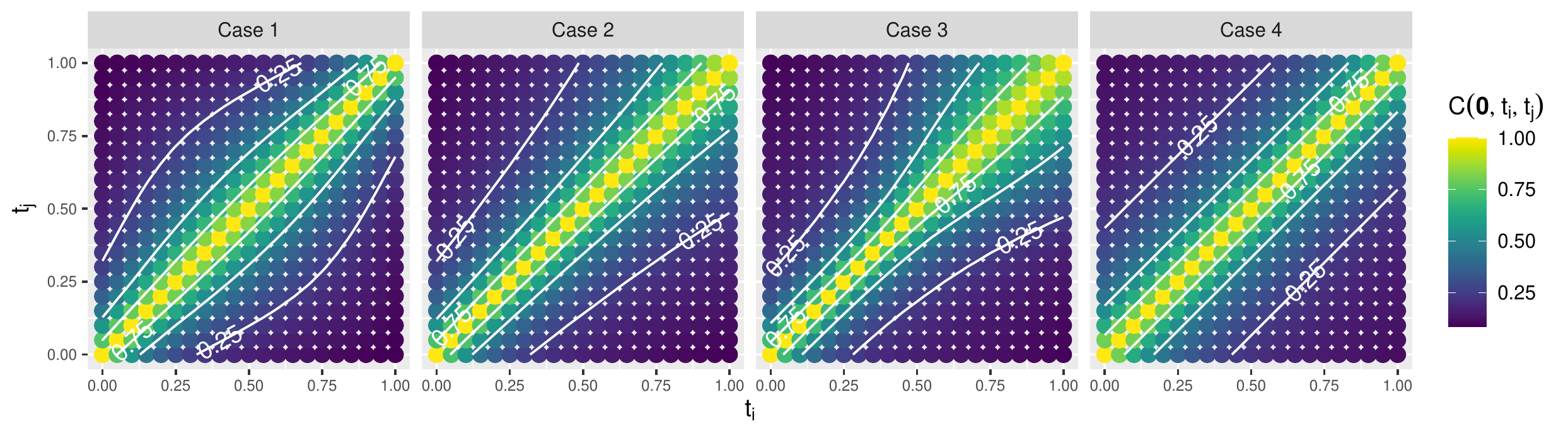}
\caption{The purely temporal covariance of the true data generating model as a function of time pairs $(t_i,t_j),\ t_i,t_j\in \mathcal{D}_t$, for all the four cases. The white lines represent the contours at levels: 0.75, 0.5 and 0.25. The purely temporal covariance in Case 1, Case 2 and Case 3 is nonstationary, as the covariance decay is slower in the middle of $\mathcal{D}_t$  for Case 1, and the covariance decay is slower at the higher end of $\mathcal{D}_t$ for Case 2 and Case 3. The purely temporal covariance in Case 4 is stationary, as the structure remain constant throughout $t$.  }
\label{fig6}
\end{figure}

\begin{figure}[h]
\centering     
\includegraphics[scale=0.55]{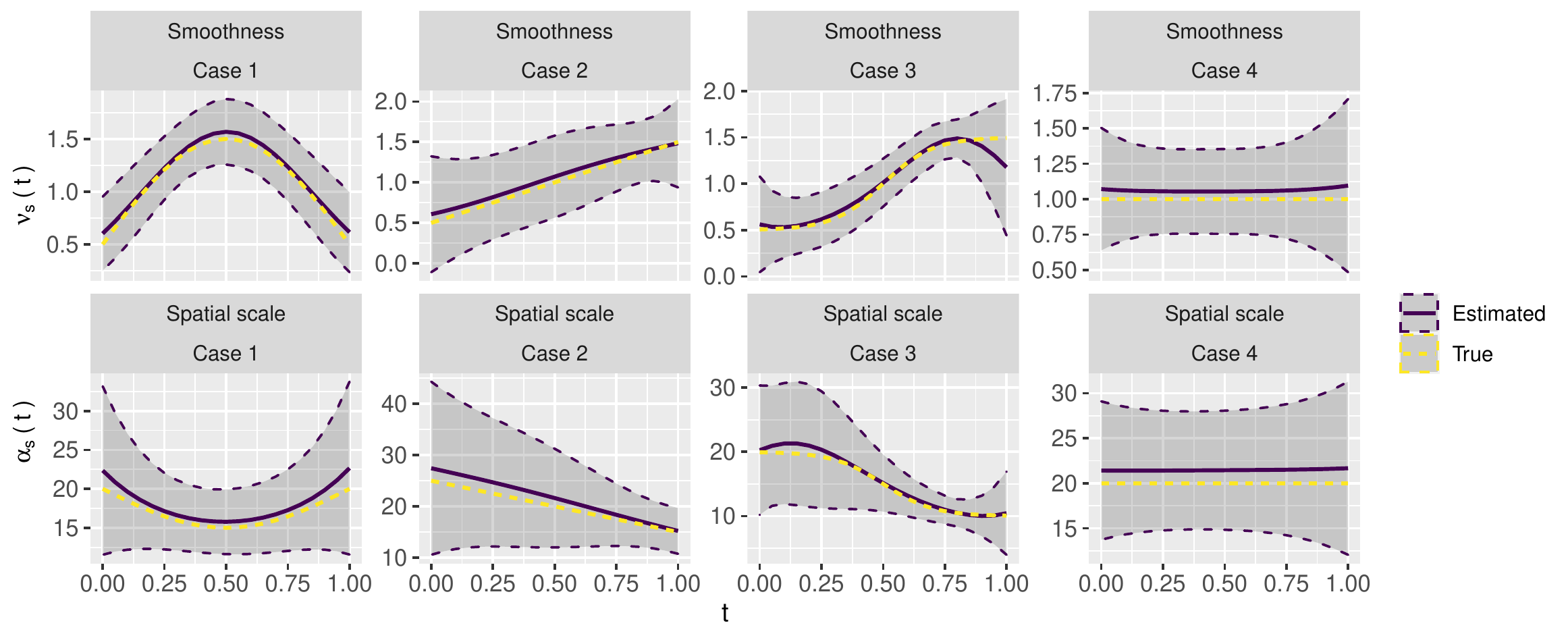}
\caption{The comparison of the pointwise average of the estimated functions $\hat{\alpha_s}(t)$ and $\hat{\nu_s}(t)$ from the candidate Tvar.M model with their corresponding true functions of the data generating Tvar.M model for all the four simulation cases. The average is taken over the 100 simulation runs. The corresponding 95\% pointwise-interval (pointwise mean $\pm$ 1.96* pointwise sd) is also shown in grey bands.}
\label{fig12}
\end{figure}

The spatial domain of interest, $\mathcal{D}_s$, is set to be $25\times25$ equally spaced grid points on a unit square, i.e., $[0,1]^2$ and temporal domain of interest, $\mathcal{D}_t$, is set as 21 equally spaced points in $[0,1]$. We simulate 100 realizations of a zero mean spatio-temporal Gaussian process $Z(\s,t),\ \s\in\mathcal{D}_s\subset\mathbb{R}^2, \ t\in\mathcal{D}_t\subset\mathbb{R}$, with Tvar.M covariance model, under four different parameter settings listed as Case 1, 2, 3 and 4 in Table \ref{tab1}. Observe that for Case 1, 2 and 3, the true functions $\alpha_s(t)$ and $\nu_s(t)$ are time varying, whereas, for Case 4, $\alpha_s(t)$ and $\nu_s(t)$ are constant; therefore, true data generating model for Case 4 is in fact Gneit.M. The true functions $\alpha_s(t)$ and $\nu_s(t)$, for $t\in\mathcal{D}_t$, are also shown in Figure \ref{fig12}. As a consequence of specified $\alpha_s(t)$ and $\nu_s(t)$, the purely spatial dependence of $Z$ varies periodically over time in Case 1, linearly over time in Case 2, nonlinearly over time in Case 3, and stays constant over time in Case 4. In terms of the  purely temporal covariance of the data generating model as shown in Figure \ref{fig6}, the specified $\alpha_s(t)$ and $\nu_s(t)$ impart nonstationarity in Case 1, Case 2 and Case 3, and stationarity in Case 4. In particular, the temporal covariance becomes stronger at the middle of $\mathcal{D}_t$ for Case 1, and at the higher end of  $\mathcal{D}_t$ for Case 2 and Case 3. An example realization of $Z$ for all the four cases can be found in the Supplementary Material.

\begin{table}
\caption{Average and standard deviation of parameter estimates over the 100 simulation runs, for the three candidate models Tvar.M, Gneit.M and Sep.M, under the four simulation cases. The second column (Parameters/Function) lists the constant parameters: $\{\sigma,a,\gamma,\beta, \delta\}$,and functional parameters: $\{\alpha_s(t),\nu_s(t)\}$ of the true data generating model Tvar.M, and the third column (True value/ True specification) provide the corresponding true values and true specification of constant and functional parameters, respectively. The true value for the parameter $a=10$, and is fixed to its true value during the estimation. }
\label{tab1}
\begin{tabular}{@{}lccccc}
\hline
&Parameter/& True value/ &\multicolumn{3}{c}{Mean (std. dev.) of the parameter estimates} \\
\cline{4-6}
Cases &Function &
True specification&
\multicolumn{1}{c}{Tvar.M} &
\multicolumn{1}{c}{Gneit.M}&
\multicolumn{1}{c}{Sep.M} \\
\hline
{Case 1} & $\sigma$ &  $1$ & $0.99$ $(0.04)$ & $0.99$ $(0.04)$ & $0.96$ $(0.04)$ \\
          
          & $\gamma$     & $0.60$ & $0.60$ $(0.02)$  & $0.61$ $(0.02)$   & $0.61$ $(0.02)$  \\
          & $\beta$     & $0.80$ & $0.75$ $(0.15)$  & $0.79$ $(0.12)$   & --  \\
          & $\delta$     & $0.10$ & $0.17$ $(0.17)$  & $0.18$ $(0.17)$   & $1.05$ $(0.15)$  \\
          & $\alpha_s(t)$     &$20+15\sin(\frac{\pi t}{20})$ & -- & $15.16$ $(2.06)$   & $25.90$ $(1.99)$ \\
          & $\nu_s(t)$      & $0.5+\sin(\frac{\pi t}{20})$ & -- & $0.94$ $(0.09)$  & $1.40$ $(0.11)$  \\[6pt]
{Case 2} & $\sigma$ &  $1$ & $0.99$ $(0.03)$ & $0.99$ $(0.03)$ & $0.96$ $(0.03)$ \\
          
          & $\gamma$     & $0.60$ & $0.60$ $(0.02)$  & $0.60$ $(0.01)$   & $0.60$ $(0.01)$  \\
          & $\beta$     & $0.80$ & $0.79$ $(0.14)$  & $0.86$ $(0.09)$   & --  \\
          & $\delta$     & $0.10$ & $0.16$ $(0.17)$  & $0.11$ $(0.08)$   & $1.03$ $(0.11)$  \\
          & $\alpha_s(t)$     & $25-10t$ & -- & $18.12$ $(1.87)$   & $32.64$ $(2.82)$ \\
          & $\nu_s(t)$      & $0.5+t$ & -- & $0.76$ $(0.07)$  & $1.23$ $(0.13)$  \\[6pt]
 {Case 3} & $\sigma$ &  $1$ & $1.00$ $(0.03)$ & $0.98$ $(0.04)$ & $0.96$ $(0.04)$ \\
          
          & $\gamma$     & $0.60$ & $0.60$ $(0.02)$  & $0.59$ $(0.01)$   & $0.59$ $(0.01)$  \\
          & $\beta$     & $0.80$ & $0.83$ $(0.13)$  & $0.90$ $(0.09)$   & --  \\
          & $\delta$     & $0.10$ & $0.10$ $(0.10)$  & $0.07$ $(0.09)$   & $1.03$ $(0.13)$  \\
          & $\alpha_s(t)$     & $20-10\frac{\exp(10t-5)}{1+\exp(10t-5)}$ & -- & $11.35$ $(1.62)$   & $18.52$ $(2.18)$ \\
          & $\nu_s(t)$      & $0.5+\frac{\exp(10t-5)}{1+\exp(10t-5)}$ & -- & $0.56$ $(0.06)$  & $0.73$ $(0.08)$  \\[6pt]         
{Case 4} & $\sigma$ &  $1$ & $0.99$ $(0.03)$ & $1.00$ $(0.03)$ & $0.97$ $(0.03)$ \\
          
          & $\gamma$     & $0.60$ & $0.60$ $(0.01)$  & $0.60$ $(0.01)$   & $0.60$ $(0.01)$  \\
          & $\beta$     & $0.80$ & $0.75$ $(0.11)$  & $0.78$ $(0.08)$   & --  \\
          & $\delta$     & $0.10$ & $0.17$ $(0.15)$  & $0.14$ $(0.12)$   & $1.00$ $(0.11)$  \\
          & $\alpha_s(t)$     & $20$& -- & $20.51$ $(2.56)$   & $38.65$ $(2.76)$ \\
          & $\nu_s(t)$      & $1$ & -- & $1.03$ $(0.10)$  & $1.83$ $(0.17)$  \\[6pt]
\hline

\end{tabular}
{\\\raggedright Note: The average estimate and standard deviation entries for $\alpha_s(t)$ and $\nu_s(t)$ are left blank under the Tvar.M since the comparison of true and estimated functional parameters $\alpha_s(t)$ and $\nu_s(t)$ under Tvar.M is shown in Figure \ref{fig12}. The entry for the average estimate and standard deviation of $\beta$ under Sep.M is left blank because $\beta=0$ for separable model.\par}
\end{table}

For comparison of interpolation and forecasting performance, we perform cross-validation, and accordingly, we split the data into a training set, a validation set for interpolation $\mathcal{V}_i$ and a validation set for forecasting $\mathcal{V}_f$. The entire spatial field at the last two time points of $\mathcal{D}_t$, i.e., at $t=\{0.95,1.00\}$, constitute $\mathcal{V}_f$. We randomly select 125 spatial locations (20\% of the total observed locations) as our validation locations for interpolation and the data at those locations for the remaining 19 time-points of $\mathcal{D}_t$, i.e., $t=\{0.00,0.05,\ldots,0.90\}$, form $\mathcal{V}_i$. All the data that remains after removing the two validation sets make our training data. For each of the four simulation cases, we fit three candidate models Tvar.M, Gneit.M and Sep.M by using the RCL estimation with $M_s=20,R_s=15,M_t=19$ and $R_t=1$, on the training data in each of the 100 simulation runs. During the estimation, the functions $\alpha_s(t)$ and $\nu_s(t)$ in the candidate model Tvar.M are specified as: $\alpha_s(t)=\exp(p^\alpha_2(t)),\ \nu_s(t)=\exp(p^\nu_2(t)),t\in\mathcal{D}_t$ for Case 1, Case 2 and Case 4, whereas for Case 3, $\alpha_s(t)=\exp(p^\alpha_3(t)),\ \nu_s(t)=\exp(p^\nu_3(t)),t\in\mathcal{D}_t$. 
Note that the specification of functions $\alpha_s(t)$ and $\nu_s(t)$ in the candidate model Tvar.M are different from that data generating model Tvar.M (see Table \ref{tab1}). Additionally, we fix $a$ to its true value, i.e., $a=10,$ in all the three candidate models during the estimation to slightly reduce the optimization burden.

Table~\ref{tab1} reports the average and standard deviation of parameter estimates over the 100 simulation runs, for all the three candidate models under the four simulation cases. The parameter estimates of $\sigma,\gamma$ and $\beta$ under the Gneit.M model are close to their respective true values in all the four cases, however,  since the Gneit.M model is misspecified for the time-varying part $\alpha_s(t)$ and $\nu_s(t)$ of the true data generating model, the respective constant estimates are not comparable to the true functions in Cases 1--3. Albeit, for Case 4 where the true $\alpha_s(t)$ and $\nu_s(t)$ are constant, the corresponding estimates under the Gneit.M model are almost equal to their true values. Among the three candidate models, Sep.M is the most extreme misspecification of the true process in all the four cases, and consequently its parameter estimates exhibit the strongest disagreement with their respective true values in all the four cases. All the parameter estimates from the candidate Tvar.M shown in Table \ref{tab1} are nearly equal to their corresponding true values, in all the four cases. In addition, the estimated $\alpha_s(t)$ and $\nu_s(t)$ from the candidate Tvar.M shown in Figure \ref{fig12} display conspicious comparability with the corresponding true functions in all the four cases. Although, it is worth noting that, in Case 3, the estimate for $\nu_s(t)$ displays increasing offset from the true values for the time period $t$ outside the training data, i.e. $t>0.90$. This points out to the fact that, outside the training temporal domain, the estimated functions $\alpha_s(t)$ and $\nu_s(t)$ should be interpreted with caution. Note that the candidate Tvar.M model is also slightly misspecified in Cases 1--3 due to its functional specification of $\alpha_s(t)$ and $\nu_s(t)$, which is different from that of the true model; however, the estimated functions $\alpha_s(t)$ and $\nu_s(t)$, in general, still recover the corresponding true functions because the specification in the candidate Tvar.M is flexible enough. Overall, these results suggest satisfactory performance of RCL method for large spatio-temporal datasets.

We now perform spatio-temporal prediction at validation space-time coordinates through kriging with the estimated three candidate covariance models to achieve cross-validation in all the four simulation cases. Specifically, we predict $Z$ at the space-time coordinates in $\mathcal{V}_i$ and $\mathcal{V}_f$ to obtain the interpolation and forecast of $Z$, respectively. Under Gaussian process framework, the predictive distribution of any unobserved ${Z}(\s^0,t^0)$ is the conditional Gaussian distribution where conditioning is over all the observed $Z(\s,t)$. Kriging provides us with prediction value $\hat{Z}(\s^0,t^0)$  and prediction variance $\hat{\sigma^2}_{\s^0,t^0}$ of the unobserved $Z(\s^0,t^0)$, which, under Gaussian process assumption, defines the predictive distribution of $Z(\s^0,t^0)$  as $\mathcal{N}\big(\hat{Z}(\s^0,t^0),\hat{\sigma^2}_{\s^0,t^0}\big)$. The predictive distribution enables us to construct $p-$prediction intervals ($p$-PI) for $Z(\s^0,t^0)$ as $\big(\hat{Q}_{\frac{1-p}{2}}(\s^0,t^0),\hat{Q}_{\frac{1+p}{2}}(\s^0,t^0)\big)$, where $\hat{Q}_p(\s^0,t^0)$ denote the $p$ quantile of $\mathcal{N}\big(\hat{Z}(\s^0,t^0),\hat{\sigma^2}_{\s^0,t^0}\big)$, and by construction, the $p-$PI includes the true value of $Z(\s^0,t^0)$ with probability $0<p<1$.  For a thorough assessment of prediction quality, the accuracy of the predicted value, prediction variance and the $p-$PI should evaluated. Accordingly, we consider the following commonly used metrics to quantify prediction performance in our cross-validation: (i) root mean squared error (RMSE), (ii) mean continuous ranked probability score (mCRPS), (iii) mean logarithmic score (mLogS) \citep{gneiting2007strictly}, (iv) Goodness statistic $(G)$ \citep{deutsch1997direct,pap_dub,pap_moy,GOOVAERTS20013}
, (iv) accuracy plot and (v) the average width plot \citep{foud,qadir2019estimation}. While the RMSE considers accuracy of only the prediction value, mCRPS and mLogS consider both the prediction value and prediction variance to assess the prediction quality. Lower values of RMSE, mCRPS and mLogS indicate superior predictions. The remaining other metrics $G$, accuracy plot and the average width plot explore the accuracy of the $p-$PI. In particular, $G\in[0,1]$ quantifies coverage accuracy of the $p-$PI, the accuracy plot visualizes the coverage accuracy through scatter plot of theoretical vs. empirical coverage of the $p-$PI over $p\in[0,1]$, and average width plot display the width of the $p-$PI as a function of $p\in[0,1]$. As a rule, higher value of $G$ and points closer to the identity line in the accuracy plot indicates better coverage; and for a fixed coverage accuracy, narrower $p$-PI is preferred. We compute all these metrics on $\mathcal{V}_i$ and $\mathcal{V}_f$ for all the three candidate models under the four simulation cases to provide a comprehensive juxtaposition of the three candidate models in terms of their predictive power. 

\begin{figure}[h]
\centering     
\includegraphics[scale=0.55]{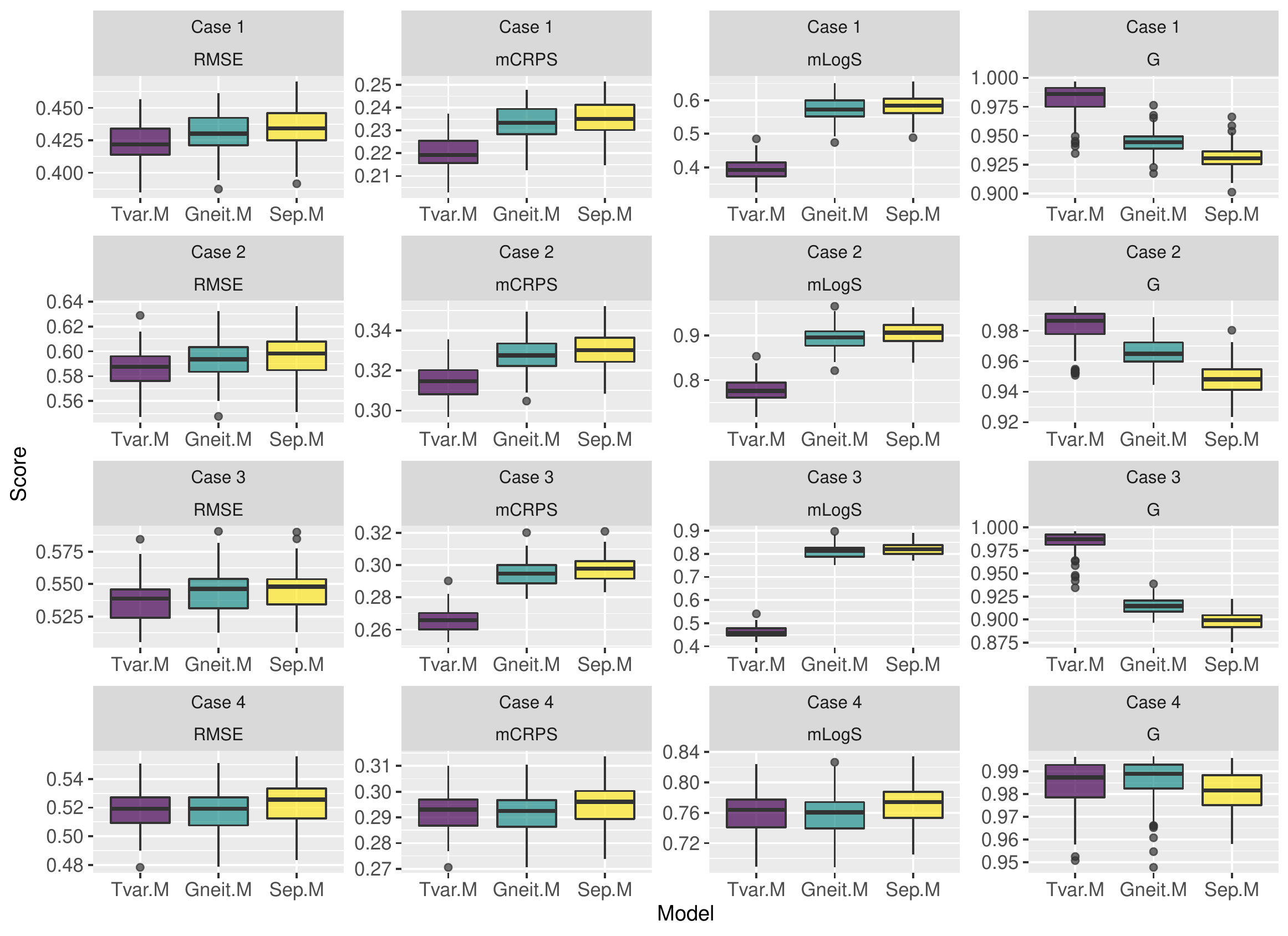}
\caption{Boxplots of 100 simulation run-based RMSE, mCRPS, mLogS and $G$, computed over the  interpolation validation set $(\mathcal{V}_i)$, for all the three candidate models under the four simulation cases. }
\label{fig8}
\end{figure}

\begin{figure}[h]
\centering     
\includegraphics[scale=0.55]{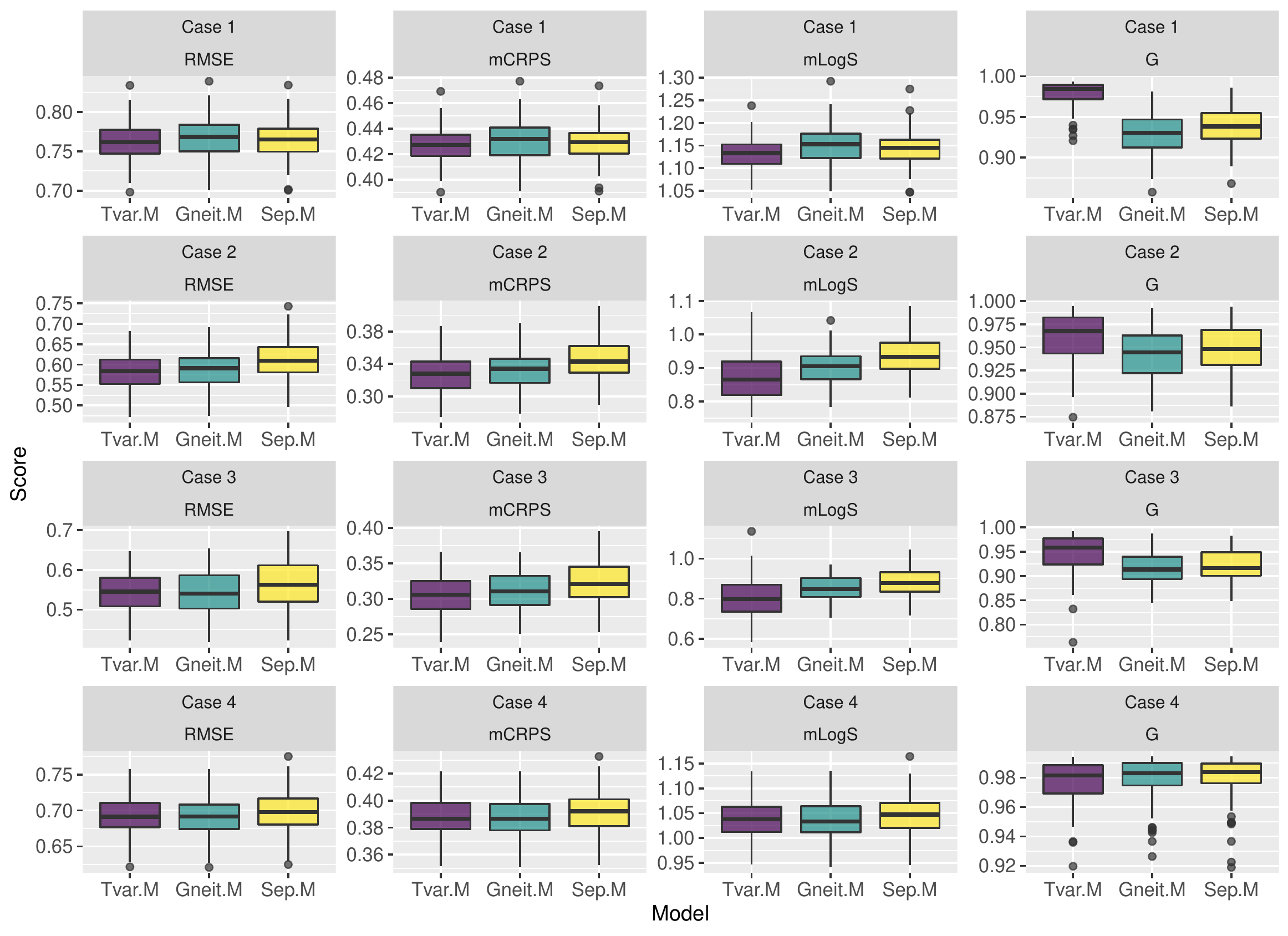}
\caption{Boxplots of 100 simulation run-based RMSE, mCRPS, mLogS and $G$, computed over the  forecasting  validation set $(\mathcal{V}_f)$, for all the three candidate models under the four simulation cases.}
\label{fig9}
\end{figure}

\begin{figure}[h]
\centering     
\includegraphics[scale=0.55]{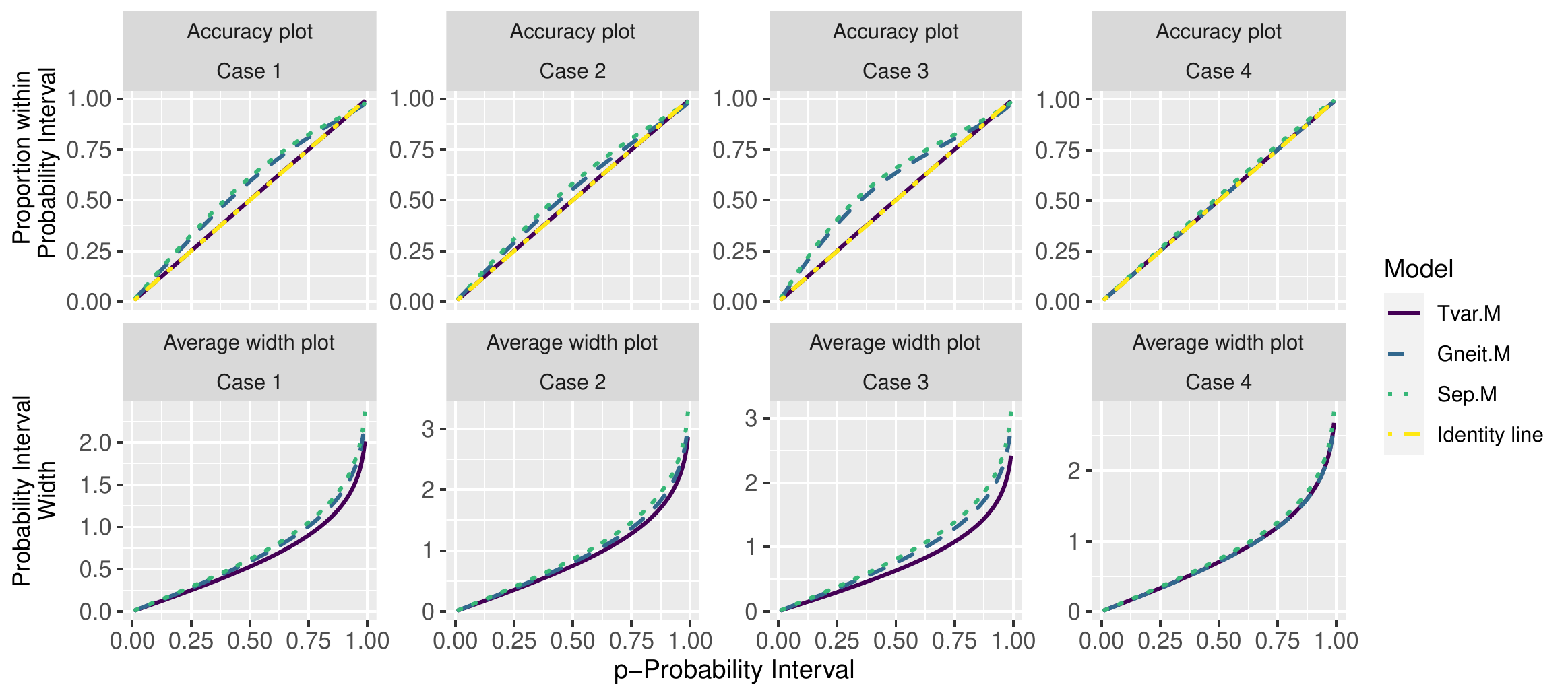}
\caption{The accuracy plot  and the average width plot  computed over the validation set for interpolation $(\mathcal{V}_i)$, for all the three candidate models under the four simulation cases. These plot represent the averaged result over the 100 simulation runs.}
\label{fig10}
\end{figure}

\begin{figure}[h]
\centering     
\includegraphics[scale=0.55]{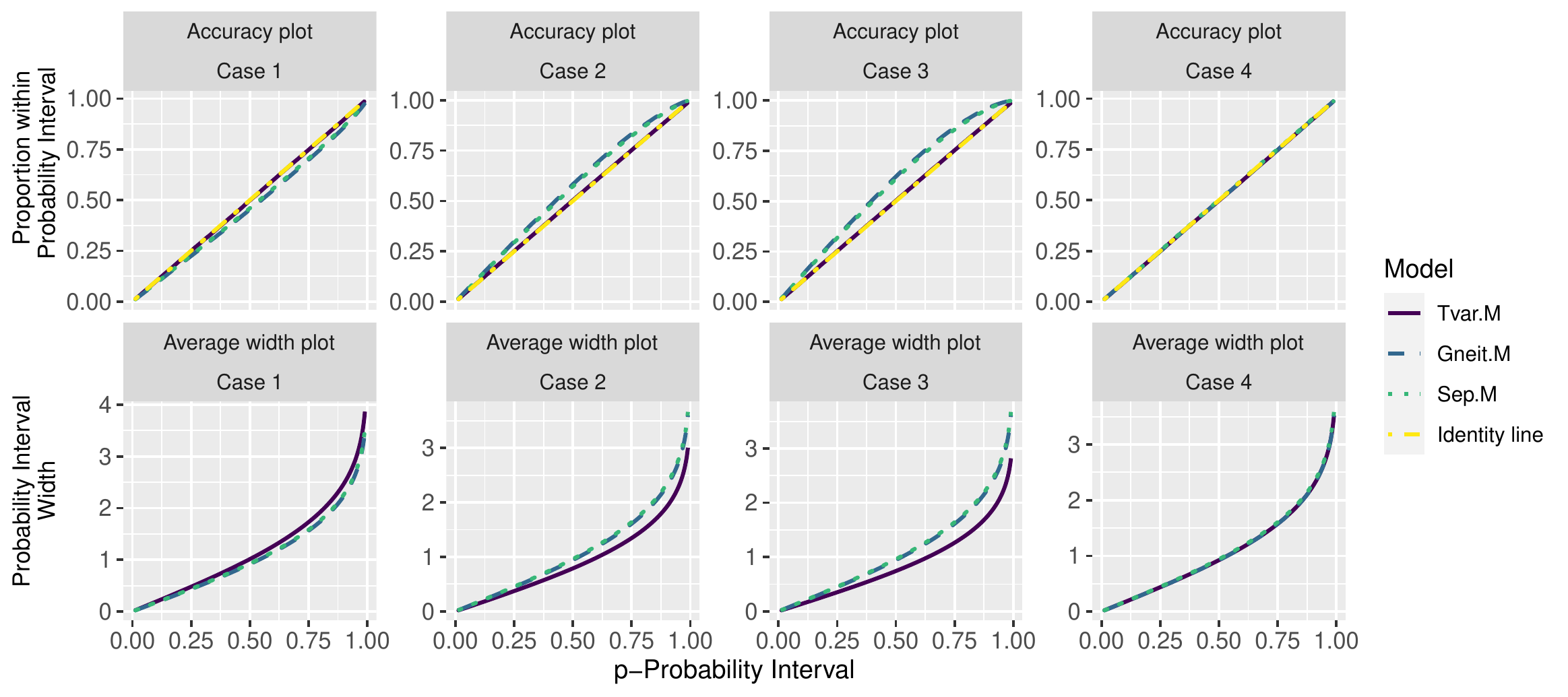}
\caption{The accuracy plot and the average width plot computed over the validation set for forecasting $(\mathcal{V}_f)$, for all the three candidate models under the four simulation cases. These plot represent the averaged result over the 100 simulation runs.}
\label{fig11}
\end{figure}

Figure \ref{fig8} shows the casewise boxplots for the RMSE, mCRPS, mLogS and $G$, computed over $\mathcal{V}_i$ for all the three candidate models  and Figure \ref{fig9} shows the same set of boxplots which are computed over $\mathcal{V}_f$ instead.  Figure \ref{fig10} shows the corresponding casewise accuracy plots and average width plots over $\mathcal{V}_i$ and Figure \ref{fig11} shows those plots for $\mathcal{V}_f$. In terms of interpolation accuracy, the Tvar.M model significantly outperforms the other two candidate models in Cases 1--3 as it produces noticeably higher $G$ and lower RMSE, mCRPS  and mLogs against Gneit.M and Sep.M (see Figure \ref{fig8}). In addition, the Tvar.M model exhibits the highest accuracy of interpolation $p-$PI (see accuracy plot in Figure \ref{fig10}) with narrowest $p$-PI width (see average width plot in Figure \ref{fig10}) among the three candidate models for Cases 1--3. These results are expected since the true underlying spatio-temporal dependence of the simulated data in Cases 1--3 is time-varying, and such dependence can be satisfactorily captured only by the Tvar.M among the three candidate models. Furthermore, between Gneit.M model and Sep.M model, the former exhibits better interpolation accuracy, although only slightly, in terms of RMSE, mCRPS and mLogS, but strongly, in terms of $G$ (see Figure \ref{fig8}) for Cases 1--3. The candidate Tvar.M model  does not exhibit any improvement in interpolation accuracy over Gneit.M in any of the assessment metrics for Case 4, which is not surprising since the the true underlying spatio-temporal dependence in Case 4 is not time-varying. Also, the interpolation accuracy of Sep.M model in Case 4 is lowest among the three candidate models, and this is attributed to the high degree of nonseparability $(\beta=0.8)$ in the simulated data. In terms of forecasting accuracy, the improvements by Tvar.M against other candidate models are clearly observed in Cases 1--3, on all the metrics (see Figure~\ref{fig9} and Figure~\ref{fig11}), except for RMSE in which the improvement are less evident. By and large, 
these results endorse the use of Tvar.M against Gneit.M and Sep.M for modeling spatio-temporal data, as the Tvar.M can potentially lead to significantly improved predictions.

\section{Data analysis}\label{dataanalysis}

In this section, we continue the data analysis from Section \ref{eda} with the same notations defined therein. As noted earlier in Section \ref{eda}, a class of time-varying spatio-temporal models is desirable for an adequate modeling of $\epsilon(\s,t)$, in accordance of which, we have developed the time-varying model \eqref{eq3}. Additionally, to handle the model estimation over the large training sample of $\epsilon(\s,t)$, which is made up of 114,310 observations through 322 spatial locations and 355 time points, we have defined RCL estimation in Section \ref{rclmethod}. We now utilize the proposed time-varying class of spatio-temporal covariance functions to model $\epsilon(\s,t)$ as a zero mean spatio-temporal Gaussian process, and then use it to perform spatio-temporal predictions through kriging. To explore the relative suitability of the proposed model for this dataset, we also consider the Gneiting-Mat\'ern class and the separable class of models in our data analysis.

\begin{figure}[h]
\centering     
\subfigure[]{\label{fig13a}\includegraphics[width=70mm]{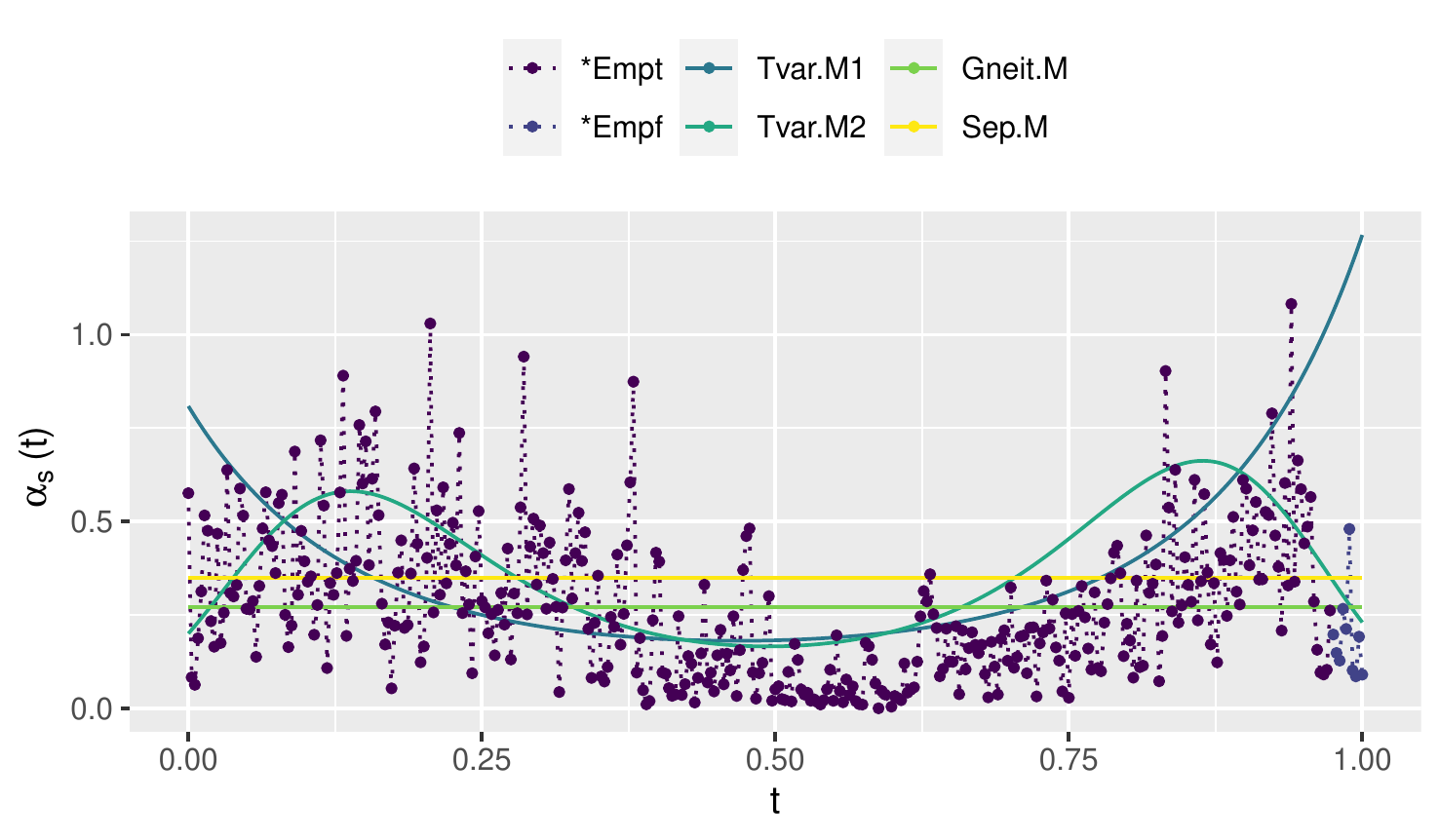}}
\subfigure[]{\label{fig13b}\includegraphics[width=70mm]{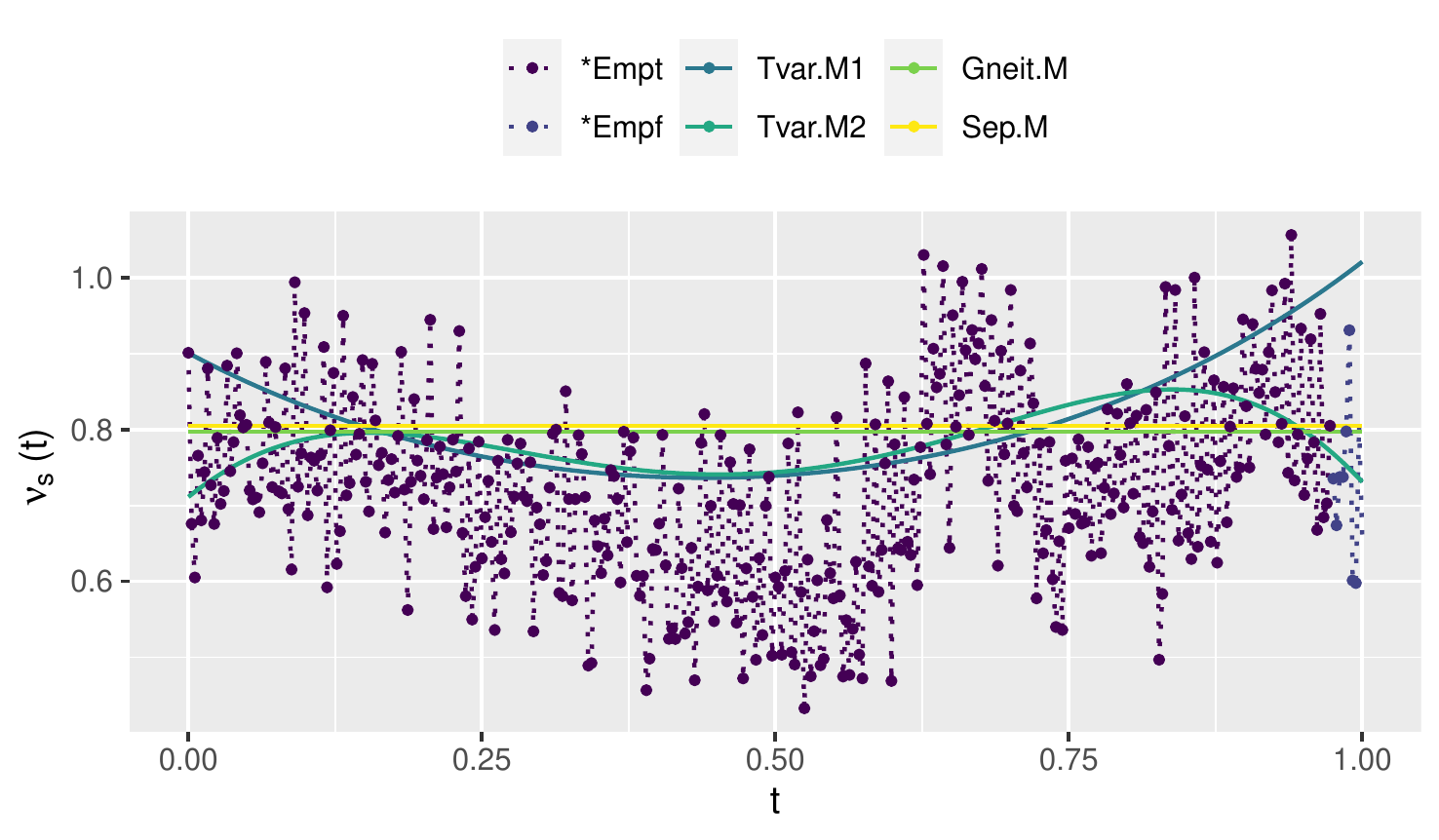}}
\caption{The estimated  (a) $\alpha_s(t)$ and (b) $\nu_s(t)$ over time $t\in[0,1]$, from  Tvar.M1, Tvar.M2, Gneit.M and Sep.M models. The estimated functions $\alpha_s(t)$ and $\nu_s(t)$ are overlayed on independent daywise maximum likelihood estimates of ${\alpha}$ and $\nu$ from $\sigma^2\text{M}(\h\mid\alpha,\nu)$, respectively. The daywise estimates corresponding to the scaled training days and forecasting days are labelled as *Empt and *Empf, respectively. Note that, here the daywise estimates are plotted on the corresponding scaled time period $t$. }
\label{fig13}
\end{figure}

\begin{table}[h]
\caption{Prediction assessment metrics  RMSE, mCRPS, mLogS and $G$, for all the four candidate models Tvar.M1, Tvar.M2, Gneit.M and Sep.M, under interpolation and forecasting}
\label{tab2}
\begin{tabular}{@{}lcccc|cccc}
\hline
Score/ &\multicolumn{4}{c}{Interpolation}  &\multicolumn{4}{|c}{Forecasting} \\
\cline{2-9}
Metric &\multicolumn{1}{c}{Tvar.M1} &\multicolumn{1}{c}{Tvar.M2} &
\multicolumn{1}{c}{Gneit.M}&
\multicolumn{1}{c}{Sep.M}&
\multicolumn{1}{|c}{Tvar.M1} &
\multicolumn{1}{c}{Tvar.M2} &
\multicolumn{1}{c}{Gneit.M}&
\multicolumn{1}{c}{Sep.M} \\
\hline
{RMSE} & $0.0474$ &   $0.0478$ &    $ 0.0482$   &      $  0.0480$ & $0.6383$   & $0.5681$   &    $0.5635 $         & $0.6138$ \\
{mCRPS} & $0.0159$  & $0.0160$  &   $ 0.0160 $   &       $0.0160$ &$0.3438 $ &$0.3053 $    &  $0.3008$         & $0.3298$\\
{mLogS} &$-2.3469$ &$-2.3624$  &   $ -2.3449$    &    $  -2.3361$ & $0.9817 $ & $0.8563 $ &   $ 0.8279 $         & $0.9452$ \\
{$G$} & $0.9496$ & $0.9462$   &    $0.9480$     &     $ 0.9480$ & $0.9683$ & $0.9839$    &   $0.9558$          & $0.9737$ \\
          
\hline

\end{tabular}

\end{table}

We consider to fit the following candidate models: Tvar.M, Gneit.M and Sep.M, on the training sample of $\epsilon(\s,t)$ by using the RCL estimation with $M_s=46,R_s=4,M_t=355$ and $R_t=1$. For the candidate model Tvar.M, we consider the following two specifications: (i) $\alpha_s(t)=\exp(p^\alpha_2(t)),\ \nu_s(t)=\exp(p^\nu_2(t))$, and (ii) $\alpha_s(t)=\exp(p^\alpha_4(t)),\ \nu_s(t)=\exp(p^\nu_4(t))$, and the resulting two variants of Tvar.M are referred to as Tvar.M1 and Tvar.M2, respectively. More specifically, we consider the following four candidate models, which are listed in the decreasing order of their flexibility as : (i) Tvar.M2, (ii) Tvar.M1, (iii) Gneit.M and (iv) Sep.M. The estimated functions $\alpha_s(t)$ and $\nu_s(t),\ t\in [0,1],$ from the four candidate models are shown in Figure \ref{fig13a} and Figure \ref{fig13b}, respectively. The time series plots in Figure \ref{fig4a} and Figure \ref{fig4b} are also included in Figure \ref{fig13a} and Figure \ref{fig13b}, respectively, after rescaling the x-axis (Day) to $t.$ Furthermore, the independent daywise maximum likelihood estimates of $\alpha$ and $\nu$ from $\sigma^2\text{M}(\h\mid \alpha,\nu)$ for day 356 to 365 (forecasting days), are also augmented in those times series in Figure \ref{fig13}. For the two variants of Tvar.M, i.e., Tvar.M1 and Tvar.M2, the estimated functions $\alpha_s(t)$ and $\nu_s(t)$ closely follow the independent daywise estimates of $\alpha$ and $\nu$, respectively, for the training days. This indicates that the candidate models Tvar.M1 and Tvar.M2 comprehend the temporally-evolving properties of the spatio-temporal process $\epsilon(\s,t)$ on the training days, which would eventually translate into better spatio-temporal interpolation. However, on the forecasting days, the candidate Tvar.M1 model seems to capture the temporally-evolving properties incorrectly, as the the estimated $\alpha_s(t)$ and $\nu_s(t)$ are completely out of sync with the daywise estimates of $\alpha$ and $\nu$, respectively. The daywise estimates exhibit a change-point for their preceeding trend near the right end of training days and since the data for forecast days are not included in the estimation, this change-point could not be captured by the estimated Tvar.M1 model. This leads to inaccuracy of Tvar.M1 for forecasting days which may affect its forecasting performance. In contrast, the more flexible variant Tvar.M2 satisfactorily accommodates those change points as estimated functions $\alpha_s(t)$ and $\nu_s(t)$ are in sync with their corresponding daywise estimates on forecasting days. This points out that Tvar.M2 is expected to perform better than Tvar.M1 in forecasting. 
The other two candidate models, i.e., Gneit.M and Sep.M, completely disregard the time-varying dependence of $\epsilon(\s,t)$ due to their theoretical limitations, and therefore, estimate constant $\alpha_s(t)$ and $\nu_s(t)$, which ignore the trends in daywise estimates. Note that, although the estimated functions $\alpha_s(t)$ and $\nu_s(t)$ from Gneit.M and Sep.M do not capture the trend of their corresponding daywise estimates, their estimated values are much closer to their corresponding daywise estimates  compared to the Tvar.M1 on forecasting days.
\begin{figure}[h]
\centering     
\includegraphics[scale=0.55]{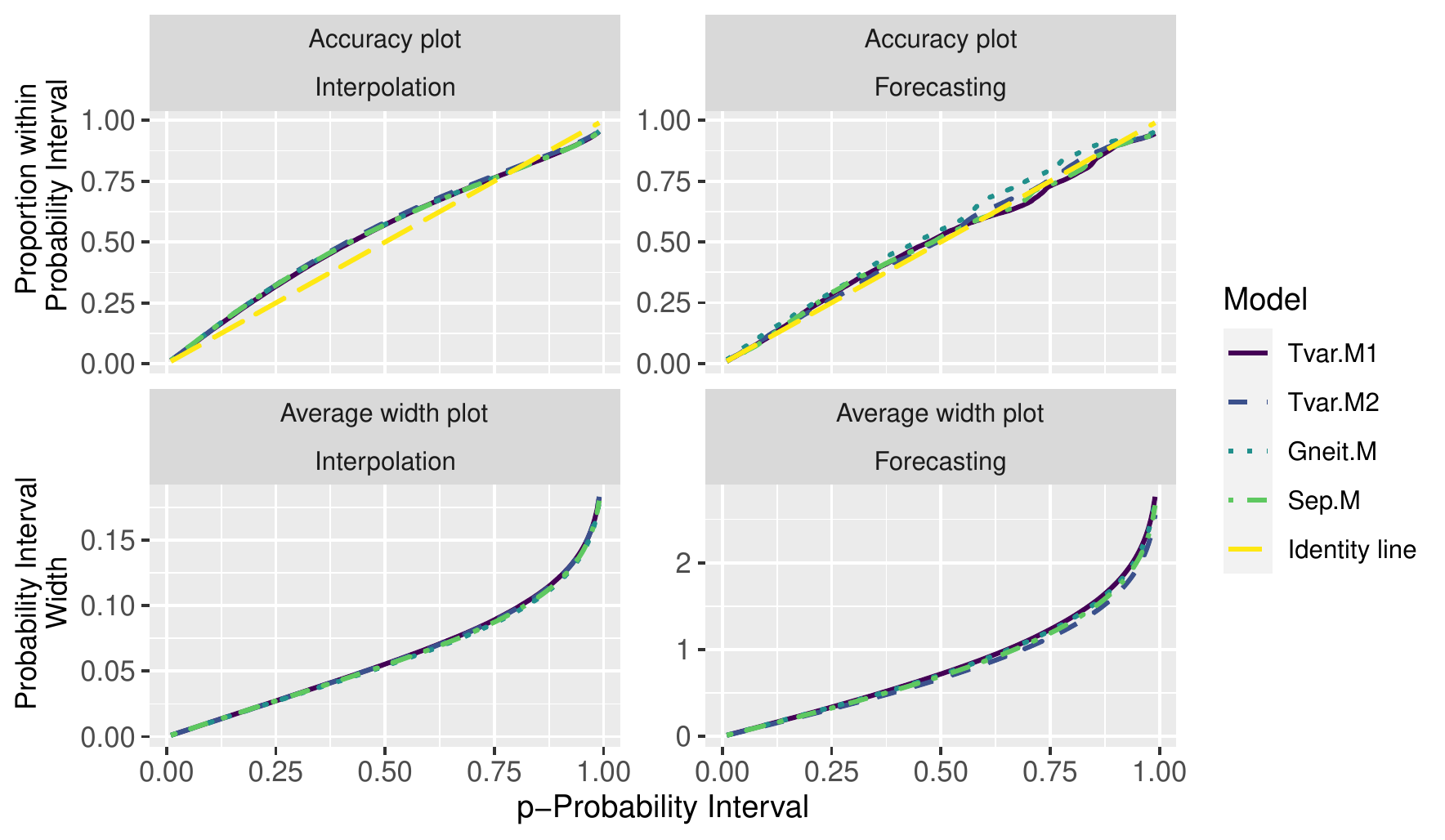}
\caption{The accuracy plot and the average width plot computed over the validation set for forecasting and interpolation, in the data analysis, for all the four candidate models.}
\label{fig14}
\end{figure}

While the estimated candidate Tvar.M variants acceptably conform to the underlying time-varying spatio-temporal dependence of  $\epsilon(\s,t)$ (see Figure \ref{fig13}), its usefulness in terms of spatio-temporal prediction still needs to be validated empirically. To this end, we now examine and compare the prediction performance of all the four candidate models for this dataset through a cross-validation study, similar to the one conducted in Section \ref{simulation}. We perform spatio-temporal prediction of $\epsilon(\s,t)$ at validation space-time coordinates, through Kriging with the four estimated candidate models. However, unlike conventional kriging where all of the observed $\epsilon(\s,t)$ is used to predict the unobserved $\epsilon(\s^0,t^0)$, we only use the observed $\epsilon(\s,t^*),\ t^*\in[t^0-6/364,t^0+6/364]$, which are in the six-time-step neighbourhood of $t^0$ for the interpolation. Additionally, for forecasting, we use all the observed $\epsilon(\s,t)$ from day 320 to day 355. This modification is done to lower the computational cost by reducing the size of the covariance matrix of the observed data which needs to be inverted in kriging. By using the predicted values, forecasting validation set, and interpolation validation set, we compute all the prediction quality assessment metrics considered in Section \ref{simulation}, for all the four estimated candidate models, under intepolation and forecasting paradigms. Table \ref{tab2} reports the RMSE, mCRPS, mLogS and $G$, under interpolation and forecasting, for all the four candidate models. The corresponding accuracy plots and the average width plots are shown in Figure \ref{fig14}. For the interpolation, the accuracy plot and average width plot does not clearly prefer any one candidate model over the other, however, the joint assessment of RMSE, mCRPS, mLogS and $G$ markedly points out the improvement by Tvar.M1 and Tvar.M2 over Gneit.M and Sep.M. Specifically, in terms of RMSE and mLogS, both Tvar.M1 and Tvar.M2 marks better interpolation accuracy than Gneit.M and Sep.M, and in terms of mCRPS and $G$, Tvar.M1 is the best candiate model for the interpolation. Concerning the forecasting performance, Tvar.M1 demonstrate the worst performance as indicated by the highest RMSE, mCRPS and mLogS, and as discussed earlier, it is expected since the time-varying features in the estimated Tvar.M1 are inaccurate for forecasting days. The forecasting performance of the Sep.M is better than Tvar.M1, however, it is significantly inferior to the Tvar.M2 and Gneit.M. Both candidate models Tvar.M2 and Gneit.M exhibit competing forecast accuracy, where on one hand Gneit.M leads on metrics like RMSE, mCRPS and mLogS, on the other hand Tvar.M2 produces significantly higher $G$ with consistently the narrowest $p$-PI. In general, these results substantiate the importance of adequate modeling of time-varying spatio-temporal dependence since the estimated Tvar.M1 and Tvar.M2 provides the best interpolation accuracy, and the estimated Tvar.M2 leads to forecast $p$-PI's with highest coverage accuracy and narrowest width.

\section{Discussion}\label{discussion}
In this article, we have developed a time-varying class of spatio-temporal models which includes the commonly used Gneiting-Mat\'ern class and separable Mat\'ern class of models as particular cases. To circumvent the challenging estimation of the proposed model over large spatio-temporal dataset, we implement a composite likelihood-based estimation method. Through our simulation study and application to the PM$_{2.5}$ data, we have established modeling advantages of our proposed class. Although, the proposed time-varying model is stationary in space and nonstationary in time, at the cost of additional complexity it can easily be made nonstationary in space as well by modifying the time varying functions $\alpha_s(t)$ and $\nu_s(t)$ to the space-time varying functions $\alpha_s(\textbf{s},t)$ and $\nu_s(\textbf{s},t)$, respectively. Specifically, if we replace $\alpha_s(t)>0$ and $\nu_s(t)>0$ with functions $\alpha_s(\textbf{s},t)>0$ and $\nu_s(\textbf{s},t)>0$, respectively, in Theorem~\ref{theorem1}, the resulting model is still valid and is nonstationary both in space and time.

The statistical analysis of the log(PM$_{2.5}$) data revealed its time-varying spatio-temporal dependence which can be probably attributed to the continuous interaction of PM$_{2.5}$ with  meteorological variables that are influenced by seasonality. Specifically, the data analysis disclose that the spatial scale and smoothness are generally lower in the middle of the year, (i.e in spring and summer), and attain peaks near the beginning and the end of the year (i.e., fall and winter). Such a spatio-temporal dependence is paramatrically modeled by the proposed class of time-varying model, and the result of which is an improvement in spatio-temporal predictions of PM$_{2.5}$.

A particular downside of the proposed model is that the time-varying functions $\alpha_s(t)$ and $\nu_s(t)$ can be sometimes misleading for the time point $t$ which is far from the training time-periods. For instance, suppose that the estimated $\nu_s(t)$ is linearly increasing in training time-period, then that would not necessarily mean that the process smoothness will be extremely high for an extremely far time point. The estimated $\nu_s(t)$ in Case 3 of simulation study illustrate this particular downside, since the estimated function $\nu_s(t)$ becomes increasingly misleading as $t$ moves away from the training time period. Additionally, the proposed model disregards the space-time asymmetry which is a commonly inherited feature in spatio-temporal datasets, therefore, introducing space-time asymmetry to the proposed class is plausible extension to this work. Lastly, it is desirable to extend the proposed class for multivariate setting, a possible direction for which is the development of multivariate analgous of Bernstein functions $\psi_{i,j}(w),\ i,j=1,\ldots,p, p\geq 1$, and using it to appropriately redefine \eqref{eq3}.

\bibliographystyle{chicago}
\bibliography{main}  
\clearpage
\begin{center}

\textbf{Supplementary Material}
\end{center}

\setcounter{section}{0}
\section{Proof of Theorem 1}
The proof of the theorem is based on considering spatio-temporal process as a multivariate spatial process with multivariate Mat\'ern covariance model \cite{multimat} and providing a valid reparameterization of a particular case of multivariate Mat\'ern model to include temporal components. Let us consider a stationary multivariate process $\textbf{Y}(\textbf{s})=\{Y_1(\s),\ldots,Y_p(\s)\}^{\text{T}},\ \s\in \mathbb{R}^d$, with mutlivariate Mat\'ern covariance model:
\begin{equation}\label{eq1as}
\text{Cov}\{{Y_i(\s),Y_j(\s+\h)}\}=\text{C}_{ij}(\h)=\rho_{ij}\sigma_i\sigma_j\text{M}(\h\mid\alpha_{ij},\nu_{ij}),
\end{equation}where the validity conditions on the model parameters $\rho_{ij},\sigma_{i},\alpha_{ij}$ and $\nu_{ij},\ i,j=1,\ldots,p,\ p\geq 1,$ are provided in Theorem 1 of \cite{multimat}. In particular, we consider the following model version derived from Corollary 1(b) of \cite{multimat}:
\begin{equation}\label{eq2as}
\text{C}_{ij}(\h)=\frac{\beta_{ij}\sigma_i\sigma_j\alpha_{ij}^d\Gamma(\frac{\nu_{i}+\nu_{j}}{2})}{\alpha_{ii}^{d/2}\alpha_{jj}^{d/2}\sqrt{\Gamma(\nu_{i})\Gamma(\nu_{j})}}\text{M}\{\h\mid\alpha_{ij},(\nu_{i}+\nu_{j})/2\}, \nu_{i},\sigma_i>0,\ i=1,\ldots,p,
\end{equation}
which is valid if: (1) $(\beta_{ij})_{i,j=1}^p$ forms a nonnegative definite matrix and (2) $(-\alpha_{ij}^{-2})_{i,j}^p$ form a conditional nonnegative definite matrix. Now, let $\beta_{ij}=1,\ i,j=1,\ldots,p$, and $\sigma_i=\sigma>0, \ i=1,\ldots,p,$ in \eqref{eq2as}, we get:

\begin{equation}\label{eq3as}
\text{C}_{ij}(\h)=\sigma^2\frac{\alpha_{ij}^d\Gamma(\frac{\nu_{i}+\nu_{j}}{2})}{\alpha_{ii}^{d/2}\alpha_{jj}^{d/2}\sqrt{\Gamma(\nu_{i})\Gamma(\nu_{j})}}\text{M}\{\h\mid\alpha_{ij},(\nu_{i}+\nu_{j})/2\}, 
\end{equation}
which is valid if $(-\alpha_{ij}^{-2})_{i,j}^p$ forms a conditional nonnegative definite matrix.

Now, let us consider a spatio-temporal process $Y(\s,t),\s\in\mathbb{R}^d,\ t\in\mathbb{R}$ such that $Y(\s,t_i)=Y_i(\s)$ for any arbitrary time-point $t_i$. Also, let $\zeta(t_i,t_j)$ be any positive valued function of time-pairs $t_i,t_j$. Corresponding adaptation of notations in \eqref{eq3as}, i.e. $\text{C}_{ij}(\h)=\text{C}(\h,t_i,t_j), \alpha_{ij}=\alpha(t_i,t_j), \nu_i=\nu_s(t_i)$, leads to the following covariance function:

\begin{equation}\label{eq4as}
\text{C}(\h,t_i,t_j)=\sigma^2\frac{\zeta(t_i,t_j)^d\Gamma\{\frac{\nu_s({t_i})+\nu_s({t_j})}{2}\}}{\zeta(t_i,t_i)^{d/2}\zeta(t_j,t_j)^{d/2}\sqrt{\Gamma\{\nu_s({t_i})\Gamma(\nu_s({t_j})\}}}\text{M}\{\h\mid\zeta(t_i,t_j),\frac{\nu_s(t_i)+\nu_s(t_j)}{2}\}, 
\end{equation}
which is valid if $\nu_s(t)>0,\ t\in\mathbb{R}$, and $-1/\zeta(t_i,t_j)^2$ forms a conditionally nonnegative definite matrix for all $t_i,t_j\in\mathbb{R}$.

Now, since $-\frac{1}{{\zeta}(t_i,t_j)^2}$ needs to form conditionally nonnegative definite matrix for all $t_i,t_j\in \mathbb{R}$, it equivalently means $\frac{1}{{\zeta}(t_i,t_j)^2}$ needs to form conditionally negative definite (cnd) matrix for all $t_i,t_j\in \mathbb{R}$. Therefore, we can use positive Bernstein functions $\psi(w)>0,w\geq0,$ to parameterize $\frac{1}{{\zeta}(t_i,t_j)^2}$. We let $\frac{1}{{\zeta}(t_i,t_j)^2}=\{\frac{\psi(|t_i-t_j|^2)}{\overline{\alpha_s}^2}+\frac{1/\alpha^2_s(t_i)+1/\alpha^2_s(t_j)}{2}-\frac{\psi(0)}{\overline{\alpha_s}^2}\},\ \overline{\alpha_s}>0, \alpha_s(t)>0, t\in\mathbb{R}$. To prove that the aforementioned parameterization is a valid parameterization, we need to show that $\{\frac{\psi(|t_i-t_j|^2)}{\overline{\alpha_s}^2}+\frac{1/\alpha^2_s(t_i)+1/\alpha^2_s(t_j)}{2}-\frac{\psi(0)}{\overline{\alpha_s}^2}\},\ \overline{\alpha_s}>0, \alpha_s(t)>0, t\in\mathbb{R}$ is conditionally negative definite.

As per \cite[~.S2]{bhatiajain}, there is a one-to-one relation between Bernstein functions and cnd functions, i.e.,``A function $\psi(\cdot)$ on $(0,\infty)$ is a Bernstein function if and only if the function $f(w)=\psi(\|w\|^2)$ is continuous and cnd on $\mathbb{R}^d$ for every $d\geq 1$. Therefore, $\psi(|t_i-t_j|^2)$ is a cnd function.

Now, to show the conditional negative definiteness of $\frac{1/\alpha^2_s(t_i)+1/\alpha^2_s(t_j)}{2}$, let $x_i\in\mathbb{C},$ such that $\sum_ix_i=0$, then,\[\sum_i\sum_jx_i\frac{1/\alpha^2_s(t_i)+1/\alpha^2_s(t_j)}{2}x_j^*\]
\[=\frac{1}{2}\sum_ix_i\frac{1}{\alpha^2_s(t_i)}\sum_j x_j^*+\frac{1}{2}\sum_j x_j^*\frac{1}{\alpha^2_s(t_j)}\sum_ix_i=0\]
Therefore, $\frac{1/\alpha^2_s(t_i)+1/\alpha^2_s(t_j)}{2}$ always forms conditionally negative definite matrix. Additionally, when $\sum_ix_i=0$, $\sum_i\sum_jx_i\psi(0)x_j^*=\psi(0)\sum_ix_i\sum_jx_j^*=0$. Now combining all the three term, we get
\[\sum_i\sum_jx_i\frac{1}{\zeta(t_i,t_j)^2}x_j^*=\sum_i\sum_jx_i\{\frac{\psi(|t_i-t_j|^2)}{\overline{\alpha_s}^2}+\frac{1/\alpha^2_s(t_i)+1/\alpha^2_s(t_j)}{2}-\frac{\psi(0)}{{\overline{\alpha_s}^2}}\}x_j^*\]
\[=\bigg[\sum_i\sum_jx_i\frac{\psi(|t_i-t_j|^2)}{{\overline{\alpha_s}^2}}x_j^*+\sum_i\sum_jx_i\frac{1/\alpha^2_s(t_i)+1/\alpha^2_s(t_j)}{2}x_j^*-\sum_i\sum_jx_i\frac{\psi(0)}{{\overline{\alpha_s}^2}}x_j^*\bigg]\]
\[\leq 0.\]

Therefore, $\frac{1}{{\zeta}(t_i,t_j)^2}=\{\frac{\psi(|t_i-t_j|^2)}{\overline{\alpha_s}^2}+\frac{1/\alpha^2_s(t_i)+1/\alpha^2_s(t_j)}{2}-\frac{\psi(0)}{\overline{\alpha_s}^2}\},\ \overline{\alpha_s}>0, \alpha_s(t)>0, t\in\mathbb{R}$ is a valid parametrization. Consequently, letting $\frac{1}{{\zeta}(t_i,t_j)^2}=\{\frac{\psi(|t_i-t_j|^2)}{\overline{\alpha_s}^2}+\frac{1/\alpha^2_s(t_i)+1/\alpha^2_s(t_j)}{2}-\frac{\psi(0)}{\overline{\alpha_s}^2}\},\ \overline{\alpha_s}>0, \alpha_s(t)>0, t\in\mathbb{R}$ in \eqref{eq4as} proves Theorem 1.
Note that, if we replace the time-varying functions $\alpha_s(t)>0$ and $\nu_s(t)>0$ with space-time varying functions $\alpha_s(\textbf{s},t)>0$ and $\nu_s(\textbf{s},t)>0$, respectively, the parameterization for $\frac{1}{{\zeta}(t_i,t_j)^2}$ would still be valid and the resulting space-time covariance would be nonstationary both in space and time.

\section{Score and Hessian for RCL}

Let $\Sigma_{\mathcal{S}_{ij},\mathcal{T}}$ and $\Sigma_{\mathcal{S},\mathcal{T}_{ij}}$ denote the covariance matrices (that depends on the parameters $\boldsymbol{\theta}$) for $\boldsymbol{X}_{\mathcal{S}_{ij},\mathcal{T}}$ and $\boldsymbol{X}_{\mathcal{S},\mathcal{T}_{ij}}$, respectively.

Under zero-mean Gaussianity, we have (ignoring the scalar terms that do not contain $\boldsymbol{\theta})$:

\begin{eqnarray}\label{eq2s}
 \ell_{RC}(\boldsymbol{\theta}\mid \boldsymbol{X}_{\mathcal{S},\mathcal{T}} )=\frac{1}{2}\times\bigg[\sum_{i=1}^{R_s}\sum_{j=1}^{M_s}\{-\frac{1}{2}\log(|\Sigma_{\mathcal{S}_{ij},\mathcal{T}}|)-\frac{1}{2}\boldsymbol{X}_{\mathcal{S}_{ij},\mathcal{T}}^T\Sigma_{\mathcal{S}_{ij},\mathcal{T}}^{-1}\boldsymbol{X}_{\mathcal{S}_{ij},\mathcal{T}}\}
\nonumber \\+ \sum_{i=1}^{R_t}\sum_{j=1}^{M_t}\{-\frac{1}{2}\log(|\Sigma_{\mathcal{S},\mathcal{T}_{ij}}|)-\frac{1}{2}\boldsymbol{X}_{\mathcal{S},\mathcal{T}_{ij}}^T\Sigma_{\mathcal{S},\mathcal{T}_{ij}}^{-1}\boldsymbol{X}_{\mathcal{S},\mathcal{T}_{ij}}\}\bigg]
\end{eqnarray}

Let $\theta_r$ denote the $r^{th}$ entry of the parameter vector $\boldsymbol{\theta}$, then we differentiate \eqref{eq2s} with respect to $\theta_r$ to obtain the score function:

\begin{eqnarray}\label{eq3s}
\frac{\partial \ell_{RC}(\boldsymbol{\theta}\mid \boldsymbol{X}_{\mathcal{S},\mathcal{T}} )}{\partial \theta_r}=\frac{1}{2}\times\bigg[\sum_{i=1}^{R_s}\sum_{j=1}^{M_s}\{-\frac{1}{2}\frac{\partial\log(|\Sigma_{\mathcal{S}_{ij},\mathcal{T}}|)}{\partial \theta_r}-\frac{1}{2}\frac{\partial \boldsymbol{X}_{\mathcal{S}_{ij},\mathcal{T}}^T\Sigma_{\mathcal{S}_{ij},\mathcal{T}}^{-1}\boldsymbol{X}_{\mathcal{S}_{ij},\mathcal{T}}}{\partial \theta_r}\}
\nonumber \\+ \sum_{i=1}^{R_t}\sum_{j=1}^{M_t}\{-\frac{1}{2}\frac{\partial \log(|\Sigma_{\mathcal{S},\mathcal{T}_{ij}}|)}{\partial \theta_r}-\frac{1}{2}\frac{\partial \boldsymbol{X}_{\mathcal{S},\mathcal{T}_{ij}}^T\Sigma_{\mathcal{S},\mathcal{T}_{ij}}^{-1}\boldsymbol{X}_{\mathcal{S},\mathcal{T}_{ij}}}{\partial \theta_r}\}\bigg]
\end{eqnarray}

In what follows, we will make use of the following formulas : (a) $\frac{\partial \log(|\Sigma|)}{\partial \theta_r}=\text{trace}(\Sigma^{-1}\frac{\partial\Sigma}{\partial \theta_r})$, (b) $\frac{\partial \textbf{Y}^T\Sigma^{-1} \textbf{Y}}{\partial\theta_r}=-\textbf{Y}^T\Sigma^{-1}\frac{\partial\Sigma}{\partial \theta_r}\Sigma^{-1}\textbf{Y}$ and (c) $\mathbb{E}(\textbf{Y}^TB\textbf{Y})=\text{trace}(B\Sigma_Y)$, where $\Sigma_Y$ is the covariance matrix for $\textbf{Y}$. Using (a) and (b) in \eqref{eq3s}, we get:

\begin{eqnarray}\label{eq4s}
\frac{\partial \ell_{RC}(\boldsymbol{\theta}\mid \boldsymbol{X}_{\mathcal{S},\mathcal{T}} )}{\partial \theta_r}=\frac{1}{2}\times\bigg[\sum_{i=1}^{R_s}\sum_{j=1}^{M_s}\{-\frac{1}{2}\text{trace}(\Sigma_{\mathcal{S}_{ij},\mathcal{T}}^{-1}\frac{\partial\Sigma_{\mathcal{S}_{ij},\mathcal{T}}}{\partial \theta_r})\nonumber \\+\frac{1}{2}\ydijt^T\edijt^{-1}\frac{\partial\edijt}{\partial \theta_r}\edijt^{-1}\ydijt\}
\nonumber \\+ \sum_{i=1}^{R_t}\sum_{j=1}^{M_t}\{-\frac{1}{2}\text{trace}(\edtij^{-1}\frac{\partial\edtij}{\partial \theta_r})+\frac{1}{2}\ydtij^T\edtij^{-1}\frac{\partial\edtij}{\partial \theta_r}\edtij^{-1}\ydtij\}\bigg]
\end{eqnarray}

Now using the formula (c) and taking expectation over both sides in \eqref{eq4s}, we get:

\begin{eqnarray}\label{eq5s}
\mathbb{E}\Big\{\frac{\partial \ell_{RC}(\boldsymbol{\theta}\mid \boldsymbol{X}_{\mathcal{S},\mathcal{T}} )}{\partial \theta_r}\Big\}=\frac{1}{2}\times\bigg[\sum_{i=1}^{R_s}\sum_{j=1}^{M_s}\{-\frac{1}{2}\text{trace}(\Sigma_{\mathcal{S}_{ij},\mathcal{T}}^{-1}\frac{\partial\Sigma_{\mathcal{S}_{ij},\mathcal{T}}}{\partial \theta_r})\nonumber \\+\frac{1}{2}\text{trace}(\edijt^{-1}\frac{\partial\edijt}{\partial \theta_r}\edijt^{-1}\edijt)\}
\nonumber \\+ \sum_{i=1}^{R_t}\sum_{j=1}^{M_t}\{-\frac{1}{2}\text{trace}(\edtij^{-1}\frac{\partial\edtij}{\partial \theta_r})+\frac{1}{2}\text{trace}(\edtij^{-1}\frac{\partial\edtij}{\partial \theta_r}\edtij^{-1}\edtij)\}\bigg]
\end{eqnarray}

\begin{eqnarray}\label{eq6s}
\mathbb{E}\Big\{\frac{\partial \ell_{RC}(\boldsymbol{\theta}\mid \boldsymbol{X}_{\mathcal{S},\mathcal{T}} )}{\partial \theta_r}\Big\}=\frac{1}{2}\times\bigg[\sum_{i=1}^{R_s}\sum_{j=1}^{M_s}\{-\frac{1}{2}\text{trace}(\Sigma_{\mathcal{S}_{ij},\mathcal{T}}^{-1}\frac{\partial\Sigma_{\mathcal{S}_{ij},\mathcal{T}}}{\partial \theta_r})+\frac{1}{2}\text{trace}(\edijt^{-1}\frac{\partial\edijt}{\partial \theta_r})\}
\nonumber \\+ \sum_{i=1}^{R_t}\sum_{j=1}^{M_t}\{-\frac{1}{2}\text{trace}(\edtij^{-1}\frac{\partial\edtij}{\partial \theta_r})+\frac{1}{2}\text{trace}(\edtij^{-1}\frac{\partial\edtij}{\partial \theta_r})\}\bigg]\nonumber \\=0
\end{eqnarray}
\textbf{Therefore, the random composite score is always an unbiased estimating function for $\boldsymbol{\theta}$}.

Now, let us consider the second derivative of \eqref{eq2s} by using the formulas (d): $d\text{trace}(AB)=\text{trace}(dA.B)+\text{trace}(A.dB)$ and (e) $\frac{\partial \Sigma^{-1}}{\partial \theta_r}=-\Sigma^{-1}\frac{\partial \Sigma}{\partial \theta_r}\Sigma^{-1}$:

\begin{eqnarray}\label{eq7s}
\frac{\partial^2 \ell_{RC}(\boldsymbol{\theta}\mid \boldsymbol{X}_{\mathcal{S},\mathcal{T}} )}{\partial \theta_r\partial \theta_s}=\frac{1}{2}\times\bigg[\sum_{i=1}^{R_s}\sum_{j=1}^{M_s}\{-\frac{1}{2}\text{trace}(\Sigma_{\mathcal{S}_{ij},\mathcal{T}}^{-1}\frac{\partial^2\Sigma_{\mathcal{S}_{ij},\mathcal{T}}}{\partial \theta_r\partial \theta_s})\nonumber \\+\frac{1}{2}\text{trace}(\edijt^{-1}\frac{\partial \Sigma_{\mathcal{S}_{ij},\mathcal{T}}}{\partial \theta_s}\edijt^{-1}\frac{\partial\Sigma_{\mathcal{S}_{ij},\mathcal{T}}}{\partial \theta_r})\nonumber \\+\frac{1}{2}\ydijt^T\edijt^{-1}\frac{\partial^2\edijt}{\partial \theta_r \partial \theta_s}\edijt^{-1}\ydijt-\nonumber \\\ydijt^T\edijt^{-1}\frac{\partial \edijt}{\partial \theta_s}\edijt^{-1}\frac{\partial \edijt}{\partial \theta_r}\edijt^{-1}\ydijt\}
\nonumber \\+ \sum_{i=1}^{R_t}\sum_{j=1}^{M_t}\{-\frac{1}{2}\text{trace}(\edtij^{-1}\frac{\partial^2\edtij}{\partial \theta_r\partial \theta_s})+\frac{1}{2}\text{trace}(\edtij^{-1}\frac{\partial \edtij}{\partial \theta_s}\edtij^{-1}\frac{\partial\edtij}{\partial \theta_r})\nonumber \\+\frac{1}{2}\ydtij^T\edtij^{-1}\frac{\partial^2\edtij}{\partial \theta_r \partial \theta_s}\edtij^{-1}\ydtij-\nonumber \\\ydtij^T\edtij^{-1}\frac{\partial \edtij}{\partial \theta_s}\edtij^{-1}\frac{\partial \edtij}{\partial \theta_r}\edtij^{-1}\ydtij\}\bigg]
\end{eqnarray}

Now taking expectation on both sides of \eqref{eq7s}, we get:

\begin{eqnarray}\label{eq8s}
\mathbb{E}\Big\{\frac{\partial^2 \ell_{RC}(\boldsymbol{\theta}\mid \boldsymbol{X}_{\mathcal{S},\mathcal{T}} )}{\partial \theta_r\partial \theta_s}\Big\}=\frac{1}{2}\times\bigg[\sum_{i=1}^{R_s}\sum_{j=1}^{M_s}\{-\frac{1}{2}\text{trace}(\Sigma_{\mathcal{S}_{ij},\mathcal{T}}^{-1}\frac{\partial^2\Sigma_{\mathcal{S}_{ij},\mathcal{T}}}{\partial \theta_r\partial \theta_s})+\nonumber \\\frac{1}{2}\text{trace}(\edijt^{-1}\frac{\partial \Sigma_{\mathcal{S}_{ij},\mathcal{T}}}{\partial \theta_s}\edijt^{-1}\frac{\partial\Sigma_{\mathcal{S}_{ij},\mathcal{T}}}{\partial \theta_r})\nonumber \\+\frac{1}{2}\text{trace}(\edijt^{-1}\frac{\partial^2\edijt}{\partial \theta_r \partial \theta_s})-\text{trace}(\edijt^{-1}\frac{\partial \edijt}{\partial \theta_s}\edijt^{-1}\frac{\partial \edijt}{\partial \theta_r})\}
\nonumber \\+ \sum_{i=1}^{R_t}\sum_{j=1}^{M_t}\{-\frac{1}{2}\text{trace}(\edtij^{-1}\frac{\partial^2\edtij}{\partial \theta_r\partial \theta_s})+\frac{1}{2}\text{trace}(\edtij^{-1}\frac{\partial \edtij}{\partial \theta_s}\edtij^{-1}\frac{\partial\edtij}{\partial \theta_r})\nonumber \\+\frac{1}{2}\text{trace}(\edtij^{-1}\frac{\partial^2\edtij}{\partial \theta_r \partial \theta_s})-\text{trace}(\edtij^{-1}\frac{\partial \edtij}{\partial \theta_s}\edtij^{-1}\frac{\partial \edtij}{\partial \theta_r})\}\bigg]
\end{eqnarray}

\begin{eqnarray}\label{eq9s}
\mathbb{E}\Big\{\frac{\partial^2 \ell_{RC}(\boldsymbol{\theta}\mid \boldsymbol{X}_{\mathcal{S},\mathcal{T}} )}{\partial \theta_r\partial \theta_s}\Big\}=\frac{1}{2}\times\bigg[\sum_{i=1}^{R_s}\sum_{j=1}^{M_s}\{-\frac{1}{2}\text{trace}(\edijt^{-1}\frac{\partial \Sigma_{\mathcal{S}_{ij},\mathcal{T}}}{\partial \theta_s}\edijt^{-1}\frac{\partial\Sigma_{\mathcal{S}_{ij},\mathcal{T}}}{\partial \theta_r})\}\nonumber \\+ \sum_{i=1}^{R_t}\sum_{j=1}^{M_t}\{-\frac{1}{2}\text{trace}(\edtij^{-1}\frac{\partial \edtij}{\partial \theta_s}\edtij^{-1}\frac{\partial\edtij}{\partial \theta_r})\}\bigg]
\end{eqnarray}

Therefore, the negative expected Hessian $H(\boldsymbol{\theta})$ is given as:

\[H(\boldsymbol{\theta})=-\mathbb{E}\Big\{\frac{\partial^2 \ell_{RC}(\boldsymbol{\theta}\mid \boldsymbol{X}_{\mathcal{S},\mathcal{T}} )}{\partial \theta_r\partial \theta_s}\Big\}\]
\begin{eqnarray}\label{eq9ab}
\nonumber=\frac{1}{4}\times\bigg[\sum_{i=1}^{R_s}\sum_{j=1}^{M_s}\{\text{trace}(\edijt^{-1}\frac{\partial \Sigma_{\mathcal{S}_{ij},\mathcal{T}}}{\partial \theta_s}\edijt^{-1}\frac{\partial\Sigma_{\mathcal{S}_{ij},\mathcal{T}}}{\partial \theta_r})\}+ \nonumber\\ \sum_{i=1}^{R_t}\sum_{j=1}^{M_t}\{\text{trace}(\edtij^{-1}\frac{\partial \edtij}{\partial \theta_s}\edtij^{-1}\frac{\partial\edtij}{\partial \theta_r})\}\bigg]
\end{eqnarray}

Typically, for the asymptotically normal estimators which result from unbiased estimating functions, the associated asymptotic covariance for the estimator has a sandwich form \citep{godambe1960} under the expanding asymptotics paradigm, and therefore: $\hat{\boldsymbol{\theta}}_{RCL}\sim N(\boldsymbol{\theta},G^{-1})$,\[G(\boldsymbol{\theta})=H(\boldsymbol{\theta})J^{-1}(\boldsymbol{\theta})H(\boldsymbol{\theta})\] where $J(\boldsymbol{\theta})=\text{var}(\frac{\partial  \ell_{RC}(\boldsymbol{\theta}\mid \boldsymbol{X}_{\mathcal{S},\mathcal{T}} )}{\partial \theta_r})$

Let us now compute the variance of the score function:

We rewrite \eqref{eq4s} by absorbing non-random terms into a constant $C$, and denoting $\ldijtr=\edijt^{-1}\frac{\partial\edijt}{\partial \theta_r}\edijt^{-1}$, and $\ldtijr=\edtij^{-1}\frac{\partial\edtij}{\partial \theta_r}\edtij^{-1}$, we get:

\begin{eqnarray}\label{eq10s}
\frac{\partial  \ell_{RC}(\boldsymbol{\theta}\mid \boldsymbol{X}_{\mathcal{S},\mathcal{T}} )}{\partial \theta_r}=\frac{1}{2}\times\bigg[\sum_{i=1}^{R_s}\sum_{j=1}^{M_s}\{\frac{1}{2}\ydijt^T\ldijtr\ydijt\}
\nonumber \\+ \sum_{i=1}^{R_t}\sum_{j=1}^{M_t}\{\frac{1}{2}\ydtij^T\ldtijr\ydtij\}\bigg]+C
\end{eqnarray}

Now, we take the variance on both sides of \eqref{eq10s} by using the formulas: (f): $\text{var}(Y^TBY)=2\text{trace}(B\Sigma_Y B\Sigma_Y)$ and (g): $\text{cov}(Y^TB_rY,Y^TB_sY)=2\text{trace}(B_r\Sigma_Y B_s\Sigma_Y)$, we get:

\begin{eqnarray}\label{eq11s}
\text{var}(\frac{\partial  \ell_{RC}(\boldsymbol{\theta}\mid \boldsymbol{X}_{\mathcal{S},\mathcal{T}} )}{\partial \theta_r})=\frac{1}{16}\times\bigg[\text{var}[\sum_{i=1}^{R_s}\sum_{j=1}^{M_s}\{\ydijt^T\ldijtr\ydijt\}]
\nonumber \\+ \text{var}[\sum_{i=1}^{R_t}\sum_{j=1}^{M_t}\{\ydtij^T\ldtijr\ydtij\}]+\nonumber \\\text{cov}(\sum_{i=1}^{R_s}\sum_{j=1}^{M_s}\{\ydijt^T\ldijtr\ydijt\},\sum_{k=1}^{R_t}\sum_{l=1}^{M_t}\{\ydtkl^T\ldtklr\ydtkl\})\bigg]
\end{eqnarray}

Let us first simplify $\text{var}[\sum_{i=1}^{R_s}\sum_{j=1}^{M_s}\{\ydijt^T\ldijtr\ydijt\}]$:

\begin{eqnarray}\label{eq12s}
\text{var}[\sum_{i=1}^{R_s}\sum_{j=1}^{M_s}\{\ydijt^T\ldijtr\ydijt\}]=\sum_{i=1}^{R_s}\text{var}(\sum_{j=1}^{M_s}\ydijt^T\ldijtr\ydijt)\nonumber\\+\sum_{l\neq m =1}^{R_s}\text{cov}(\sum_{j=1}^{M_s}\ydljt^T\ldljtr\ydljt,\sum_{n=1}^{M_s}\ydmnt^T\ldmntr\ydmnt)
\end{eqnarray}

\begin{eqnarray}\label{eq13s}
\text{var}[\sum_{i=1}^{R_s}\sum_{j=1}^{M_s}\{\ydijt^T\ldijtr\ydijt\}]=\sum_{i=1}^{R_s}\sum_{j=1}^{M_s}\text{var}(\ydijt^T\ldijtr\ydijt)\nonumber\\+\sum_{i=1}^{R_s}\sum_{j\neq j'}^{M_s}\text{cov}(\ydijt^T\ldijtr\ydijt,\ydijdt^T\ldijdtr\ydijdt)\nonumber\\+\sum_{l\neq m =1}^{R_s}\sum_{j=1}^{M_s}\sum_{n=1}^{M_s}\text{cov}(\ydljt^T\ldljtr\ydljt,\ydmnt^T\ldmntr\ydmnt)
\end{eqnarray}

\begin{eqnarray}\label{eq14s}
\text{var}[\sum_{i=1}^{R_s}\sum_{j=1}^{M_s}\{\ydijt^T\ldijtr\ydijt\}]=2\sum_{i=1}^{R_s}\sum_{j=1}^{M_s}\text{trace}(\ldijtr\edijt\ldijtr)\nonumber\\+2\sum_{i=1}^{R_s}\sum_{j\neq j'}^{M_s}\text{trace}(\overline{\ldijtr}\Sigma_{\mathcal{S}_{ij},\mathcal{S}_{ij'},\mathcal{T}}\underline{\ldijdtr}\Sigma_{\mathcal{S}_{ij},\mathcal{S}_{ij'},\mathcal{T}})\nonumber\\+2\sum_{l\neq m =1}^{R_s}\sum_{j=1}^{M_s}\sum_{n=1}^{M_s}\text{trace}(\overline{\ldljtr}\Sigma_{\mathcal{S}_{ij},\mathcal{S}_{mn},\mathcal{T}}\underline{\ldmntr}\Sigma_{\mathcal{S}_{ij},\mathcal{S}_{mn},\mathcal{T}}),
\end{eqnarray}
where $\Sigma_{\mathcal{S}_{ij},\mathcal{S}_{mn},\mathcal{T}}$ is the covariance matrix for $(\ydijt^T,\ydijdt^T)^T$, \[\overline{\ldijtr}=\begin{bmatrix}
\ldijtr & 0 \\
0 & 0 
\end{bmatrix},\quad \underline{\ldijdtr}=\begin{bmatrix}
0 & 0 \\
0 & \ldijdtr 
\end{bmatrix}\]

Similarly, we get:
\begin{eqnarray}\label{eq15s}
\text{var}[\sum_{i=1}^{R_t}\sum_{j=1}^{M_t}\{\ydtij^T\ldtijr\ydtij\}]= 2\sum_{i=1}^{R_t}\sum_{j=1}^{M_t}\text{trace}(\ldtijr\edtij\ldtijr)\nonumber\\+2\sum_{i=1}^{R_t}\sum_{j\neq j'}^{M_t}\text{trace}(\overline{\ldtijr}\Sigma_{\mathcal{S},\mathcal{T}_{ij},\mathcal{T}_{ij'}}\underline{\ldtijr}\Sigma_{\mathcal{S},\mathcal{T}_{ij},\mathcal{T}_{ij'}})\nonumber\\+2\sum_{l\neq m =1}^{R_t}\sum_{j=1}^{M_t}\sum_{n=1}^{M_t}\text{trace}(\overline{\ldtljr}\Sigma_{\mathcal{S},\mathcal{T}_{ij},\mathcal{T}_{mn}}\underline{\ldtmnr}\Sigma_{\mathcal{S},\mathcal{S}_{ij},\mathcal{T}_{mn}})
\end{eqnarray}

and

\begin{eqnarray}\label{eq16s}
     \sum_{i=1}^{R_s}\sum_{j=1}^{M_s}\sum_{k=1}^{R_t}\sum_{l=1}^{M_t}\text{cov}(\ydijt^T\ldijtr\ydijt,\ydtkl^T\ldtklr\ydtkl)=\nonumber\\ 2\sum_{i=1}^{R_s}\sum_{j=1}^{M_s}\sum_{k=1}^{R_t}\sum_{l=1}^{M_t} \text{trace}(\overline{\ldijtr}\Sigma_{\mathcal{S}_{ij},\mathcal{T},\mathcal{S},\mathcal{T}_{kl}}\underline{\ldtklr}\Sigma_{\mathcal{S}_{ij},\mathcal{T},\mathcal{S},\mathcal{T}_{kl}}), 
\end{eqnarray}
where $\Sigma_{\mathcal{S}_{ij},\mathcal{T},\mathcal{S},\mathcal{T}_{kl}}$ is the covariance matrix of $(\ydijt^T,\ydtkl^T)^T$

Therefore, by using \eqref{eq14s},\eqref{eq15s} and \eqref{eq16s}, we can obtain the diagonal entries of $J(\boldsymbol{\theta})$: 

\begin{eqnarray}\label{eq17s}
    \text{var}(\frac{\partial \ell_{RC}(\boldsymbol{\theta}\mid \boldsymbol{X}_{\mathcal{S},\mathcal{T}} )}{\partial \theta_r})=\frac{1}{8}\times\Bigg[\sum_{i=1}^{R_s}\sum_{j=1}^{M_s}\text{trace}(\ldijtr\edijt\ldijtr)\nonumber\\+\sum_{i=1}^{R_s}\sum_{j\neq j'}^{M_s}\text{trace}(\overline{\ldijtr}\Sigma_{\mathcal{S}_{ij},\mathcal{S}_{ij'},\mathcal{T}}\underline{\ldijdtr}\Sigma_{\mathcal{S}_{ij},\mathcal{S}_{ij'},\mathcal{T}})\nonumber\\+\sum_{l\neq m =1}^{R_s}\sum_{j=1}^{M_s}\sum_{n=1}^{M_s}\text{trace}(\overline{\ldljtr}\Sigma_{\mathcal{S}_{ij},\mathcal{S}_{mn},\mathcal{T}}\underline{\ldmntr}\Sigma_{\mathcal{S}_{ij},\mathcal{S}_{mn},\mathcal{T}})\nonumber\\+\sum_{i=1}^{R_t}\sum_{j=1}^{M_t}\text{trace}(\ldtijr\edtij\ldtijr)\nonumber\\+\sum_{i=1}^{R_t}\sum_{j\neq j'}^{M_t}\text{trace}(\overline{\ldtijr}\Sigma_{\mathcal{S},\mathcal{T}_{ij},\mathcal{T}_{ij'}}\underline{\ldtijr}\Sigma_{\mathcal{S},\mathcal{T}_{ij},\mathcal{T}_{ij'}})\nonumber\\+\sum_{l\neq m =1}^{R_t}\sum_{j=1}^{M_t}\sum_{n=1}^{M_t}\text{trace}(\overline{\ldtljr}\Sigma_{\mathcal{S},\mathcal{T}_{ij},\mathcal{T}_{mn}}\underline{\ldtmnr}\Sigma_{\mathcal{S},\mathcal{S}_{ij},\mathcal{T}_{mn}})\nonumber\\+\sum_{i=1}^{R_s}\sum_{j=1}^{M_s}\sum_{k=1}^{R_t}\sum_{l=1}^{M_t} \text{trace}(\overline{\ldijtr}\Sigma_{\mathcal{S}_{ij},\mathcal{T},\mathcal{S},\mathcal{T}_{kl}}\underline{\ldtklr}\Sigma_{\mathcal{S}_{ij},\mathcal{T},\mathcal{S},\mathcal{T}_{kl}})\Bigg]. 
\end{eqnarray}
Similarly, we can obtain the off-diagonal entries of $J(\boldsymbol{\theta})$ and plug it in the formula of the $G(\boldsymbol{\theta})$ to obtain the variance of the parameter estimates $\hat{\boldsymbol{\theta}}_{RCL}$.

\section{Additional Figures from the Simulation Study}
In this section, we present some additional figure from the simulation study presented in the main paper.

\begin{figure}[h]
\centering     
\includegraphics[scale=0.60]{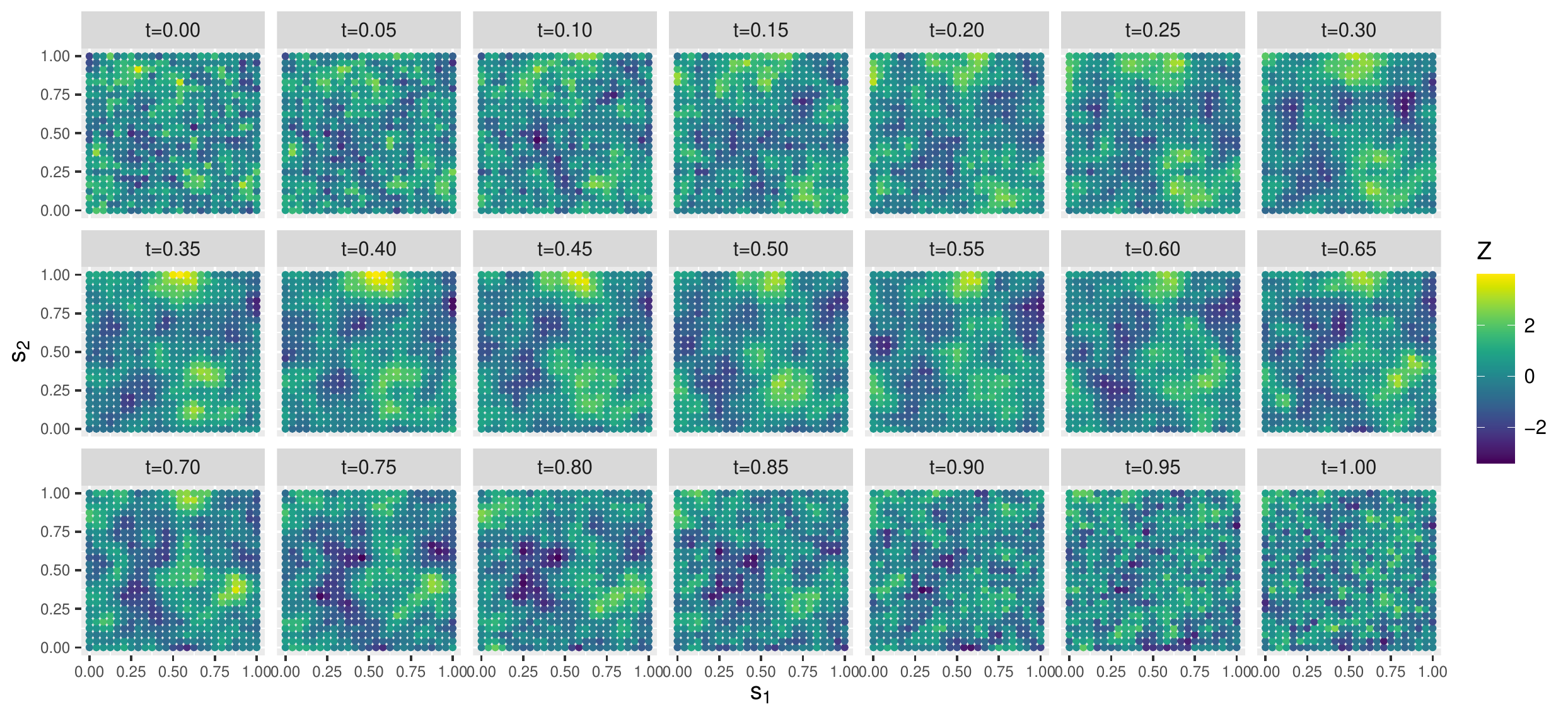}
\caption{An example of a simulated realization of $Z$ from Case1.}
\label{fig1s}
\end{figure}

\begin{figure}[h]
\centering     
\includegraphics[scale=0.60]{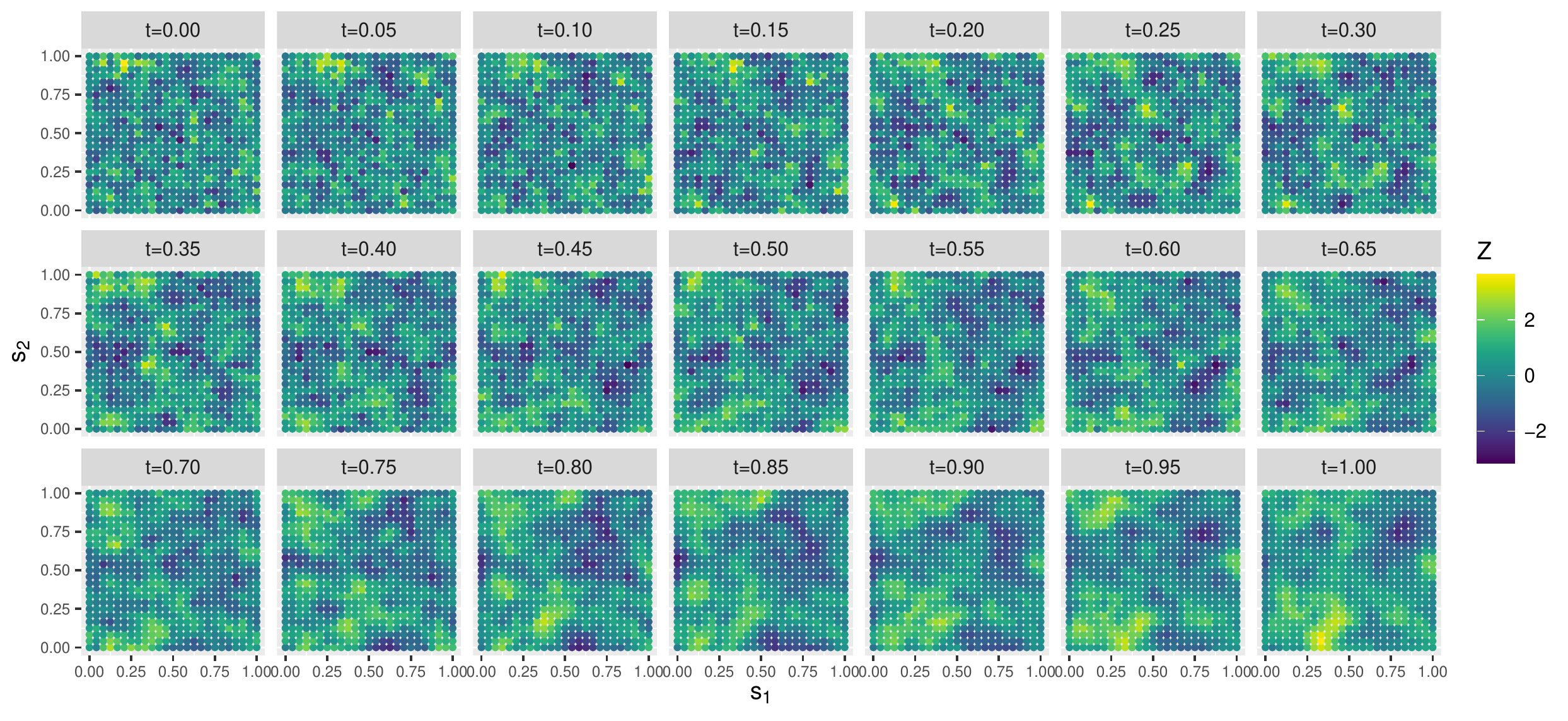}
\caption{An example of a simulated realization of $Z$ from Case2.}
\label{fig1s}
\end{figure}

\begin{figure}
\centering     
\includegraphics[scale=0.60]{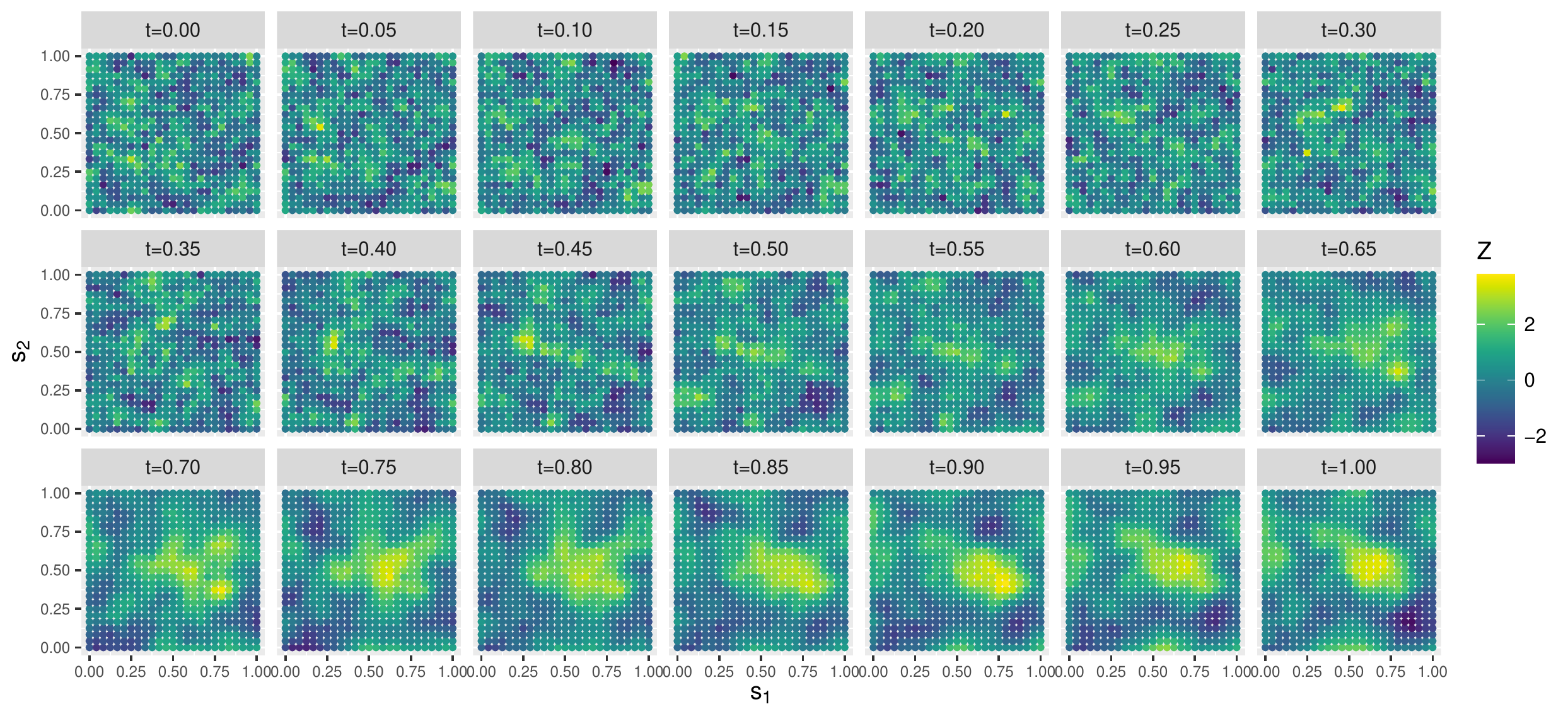}
\caption{An example of a simulated realization of $Z$ from Case3.}
\label{fig2s}
\end{figure}

\begin{figure}
\centering     
\includegraphics[scale=0.60]{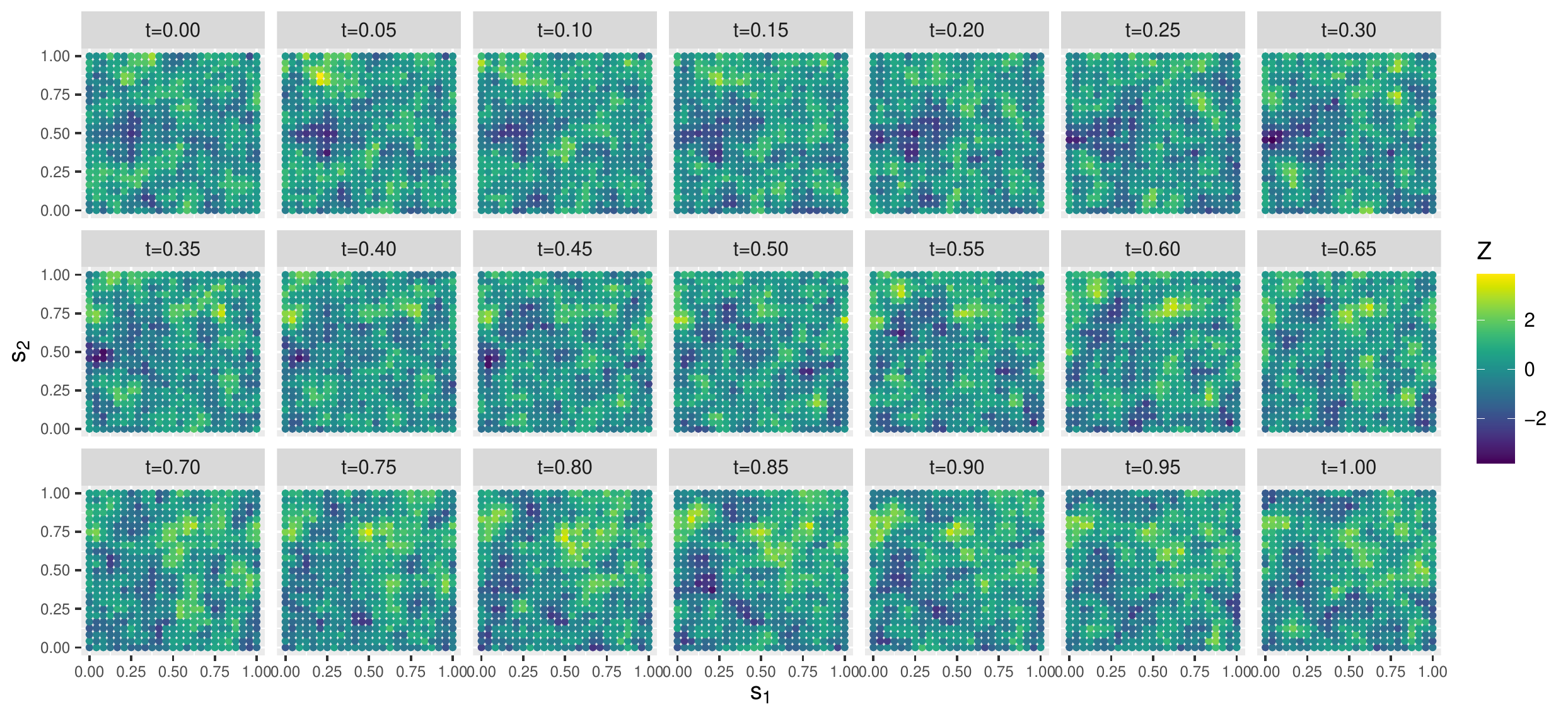}
\caption{An example of a simulated realization of $Z$ from Case4.}
\label{fig2s}
\end{figure}


\end{document}